\PassOptionsToPackage{dvipsnames}{xcolor}

\documentclass[10pt,journal]{IEEEtran}

\usepackage{tikz}
\usepackage[most]{tcolorbox}
\tcbuselibrary{listings,skins,breakable}

\usepackage{siunitx}
\usepackage{colortbl}

\usetikzlibrary{positioning,arrows.meta,shapes,fit,calc,backgrounds,patterns,decorations.pathmorphing}

\usepackage{pgfplots}
\usepackage{pgfplotstable}
\pgfplotsset{compat=1.18}
\usepgfplotslibrary{statistics}

\newtcolorbox{rqbox}{
    colback=gray!10,
    colframe=black,
    boxrule=0.5pt,
    arc=0pt,
    left=2pt, right=2pt, top=2pt, bottom=2pt,
    boxsep=2pt
}

\newtcolorbox{rqanswerbox}{
    colback=gray!10,
    colframe=black,
    boxrule=0.5pt,
    arc=0pt,
    left=2pt, right=2pt, top=2pt, bottom=2pt,
    boxsep=2pt,
    breakable
}

\usepackage{xspace}

\newcommand{\eg}{\emph{e.g.}\xspace}
\newcommand{\ie}{\emph{i.e.}\xspace}

\newcommand{\fixme}[1]{\textbf{\textcolor{red}{[Fixme: #1]}}}

\newcommand{\sysname}{ARCADE}

\usepackage{amsmath}
\usepackage{booktabs}
\usepackage{tabularx}
\usepackage{subcaption}
\usepackage{enumitem}
\usepackage{amsthm}
\usepackage{amsfonts}
\usepackage{multirow}
\usepackage{pifont}
\newcommand{\cmark}{\ding{51}}
\newcommand{\xmark}{\ding{55}}

\usepackage{soul}
\usepackage{balance}

\newtheoremstyle{definitionstyle}
  {6pt}{6pt}{}{}{\bfseries}{.}{ }{}
\theoremstyle{definitionstyle}
\newtheorem{definition}{Definition}
\newtheorem{property}{Property}

\graphicspath{{figures/}}
\definecolor{rowgray}{gray}{0.95}
\definecolor{headergray}{gray}{0.90}
\definecolor{codeblack}{rgb}{0.13, 0.13, 0.13}
\definecolor{codegray}{rgb}{0.5, 0.5, 0.5}
\definecolor{codecomment}{rgb}{0.4, 0.4, 0.4}
\definecolor{codestring}{rgb}{0.13, 0.54, 0.13}
\definecolor{codekeyword}{rgb}{0.0, 0.0, 0.8}
\definecolor{codebackground}{rgb}{0.95, 0.95, 0.95}
\definecolor{codeframe}{rgb}{0.7, 0.7, 0.7}

\newtcblisting[auto counter]{prolisting}[3][]{
    enhanced,
    listing engine=listings,
    colback=codebackground,
    colframe=codeframe,
    boxrule=0.5pt,
    arc=0pt,
    left=5pt, right=5pt, top=5pt, bottom=5pt,
    listing only,
    breakable,
    after upper={\vskip3pt\relax\centering\small\textbf{Listing \thetcbcounter:} #2},
    label={#3},
    listing options={
        basicstyle=\ttfamily\scriptsize\color{codeblack},
        keywordstyle=\color{codekeyword}\bfseries,
        commentstyle=\color{codecomment}\itshape,
        stringstyle=\color{codestring},
        numbers=left,
        numberstyle=\tiny\color{codegray},
        numbersep=8pt,
        tabsize=2,
        breaklines=true,
        showstringspaces=false,
        language=C++,
        morekeywords={grant, deny, ignore},
        aboveskip=0pt,
        belowskip=0pt,
        frame=none,
    },
    #1
}

\lstdefinelanguage{Cedar}{
  keywords={permit, forbid, principal, action, resource, when, unless, in, if, then, else, true, false, context},
  keywordstyle=\color{blue}\bfseries,
  ndkeywords={User, Action, Table, Database, Column},
  ndkeywordstyle=\color{purple},
  sensitive=true,
  comment=[l]{//},
  morecomment=[s]{/*}{*/},
  morestring=[b]",
  stringstyle=\color{red},
}

\lstdefinelanguage{JSON}{
  basicstyle=\ttfamily\scriptsize,
  stringstyle=\color{red},
  identifierstyle=\color{black},
  keywordstyle=\color{blue},
  keywordstyle=[2]\color{magenta},
  keywordstyle=[3]\color{cyan},
  keywords={},
  morekeywords=[2]{principal, action, resource, context, time, day, date, ip},
  morekeywords=[3]{__extn, fn, arg},
  morestring=[b]",
  morestring=[b]',
  comment=[l]{//},
  commentstyle=\color{gray}\ttfamily,
  numbers=left,
  numberstyle=\tiny\color{gray},
  breaklines=true,
  showstringspaces=false,
  frame=single,
  rulecolor=\color{black},
  backgroundcolor=\color{white},
}

\usepackage[maxbibnames=2, doi=false]{biblatex} 
\AtEveryBibitem{%
  \clearfield{urlyear}\clearfield{urlmonth}\clearfield{urlday}%
  \clearfield{note}\clearfield{abstract}\clearlist{keywords}%
}

\begin{document}

\hyphenation{au-thor-i-za-tion sub-stan-tial ar-chi-tec-ture}
\title{\sysname{}: Retrofitting Databases with ABAC using External Authorization Engines}



 \author{Karthikeya S M Yelisetty, Shamik~Sural and~Santosh~Nagarakatte~\IEEEmembership{}  

\thanks{Karthikeya S M Yelisetty (yk553@kgpian.iitkgp.ac.in) and Shamik Sural (shamik@cse.iitkgp.ac.in) are with the Department of Computer Science and Engineering, Indian Institute of Technology, Kharagpur, India. Santosh~Nagarakatte (santosh.nagarakatte@cs.rutgers.edu) is with Rutgers University, USA}
}

\markboth{}
{Yelisetty \MakeLowercase{\textit{et al.}}: \sysname{}: Retrofitting Databases with ABAC using External Authorization Engines}

\maketitle

\maketitle
\begin{abstract}
This paper proposes a methodology for decoupling the authorization logic of a relational database from its core data management engine, enabling specification of advanced access control policies with minimal changes to the code base. Our decoupled authorization architecture, which we call \sysname{}, externalizes access request resolution to a formally verifiable policy engine (\eg, Cedar), while the database itself enforces those decisions. We showcase the capabilities of \sysname{} on two widely popular open-source databases, namely MySQL and PostgreSQL. Purpose built lightweight plugins add an Attribute-based Access Control (ABAC) layer that augments the native discretionary and role-based access control capabilities of these databases. \sysname{} supports highly expressive policies using attributes of subjects, objects, and the environment, and the use of Cedar enables easy auditing and verification of the specified policies. It follows a design principle that mitigates potential risks in segregating authorization from the database core through low footprint, network isolation, and  a policy composition methodology that does not negate any native authorization. Experiments on TPC-C, TPC-B, and a custom workload show that \sysname{} incurs negligible performance overhead, for instance, less than 3.9\% throughput penalty for MySQL (1.3\% for PostgreSQL) on industry standard TPC-C benchmark.
\end{abstract}

\begin{IEEEkeywords}
    Attribute-based access control, Database authorization, Cedar, External authorization
\end{IEEEkeywords}


\section{Introduction}
\label{sec:intro}

In recent years, incidents of database breaches have resulted in unauthorized access to sensitive data for millions of users. An investigation report by Verizon showed that a leading cause of these breaches is credential compromise and privilege abuse~\cite{verizon2024dbir}. Yet, the authorization models protecting such databases, which are predominantly relational, have
remained practically unchanged over several decades~\cite{griffiths1976authorization}.
Mainstream open-source relational databases still primarily rely on discretionary and
role-based access control (DAC/RBAC) models~\cite{eisenbergSQL1999Formerly1999}.
These models were originally designed for an era when trusted operators
and single-tenant deployments used to be more prevalent, and are not capable of expressing policies that depend on specific context like
time of day, client location, or risk signals, which modern security
practices demand.

Attribute-based access control (ABAC)~\cite{hu2014guide} addresses this
gap by evaluating access requests against a set of rules (collectively called a policy)  constructed from a
rich set of attributes of the principal, resource, action, and
environment. Consider a health-care system where physicians can access
patient records only within their specialty, during working hours,
and from the hospital intranet. Encoding such a policy in a relational
database with DAC/RBAC models (using \texttt{GRANT} and \texttt{CREATE ROLE} statements natively
available in SQL) results in the creation of a large number of roles.
For example, a specific role has to be created for
physicians in the cardiology department working at specific hours. Administrators who avoid this proliferation settle for a coarser role, granting more access than the task requires, thus violating the principle of least
privilege~\cite{saltzer1975protection}.
Another alternative is to
implement access control in application code, which, however, would scatter
security logic across services, exposing the database to
direct access paths.

The database community has made some, though limited, explorations towards incorporating
ABAC into database authorization. These isolated attempts make
suboptimal design choices that require use of external gateways (\ie,
proxy-based enforcement) or significant modifications to the entire
database code. A gateway is meant to intercept connections, but it can potentially be bypassed by anyone
having direct access to the database, thereby undermining the overall security.
Alternatively, modifying the
database engines to support ABAC models involves invasive changes to several thousand lines in the
most sensitive code paths, requiring extensive re-integration efforts on every database
release. Hence, these approaches can seldom be used in production.

It is in this context that we propose \sysname{} (\textbf{A}ttribute-based \textbf{R}etrofit for 
\textbf{C}ontext-aware \textbf{A}uthorization in \textbf{D}atabase \textbf{E}ngines) -- an alternative authorization architecture requiring minimal changes to a database while enhancing it with support for ABAC. Almost all mainstream relational databases provide extension APIs (\eg MySQL
plugins~\cite{mysql8-plugins} and PostgreSQL hooks~\cite{H4Extensions2025}), for 
enhancing their functionality
without modifying the database core. To achieve our goal, 
we identify and develop minimal shim layers: an authorization plugin 
interface for MySQL and a custom authorization hook for PostgreSQL.
The shim layer intercepts authorization checks only when native
DAC and RBAC deny access, delegating the final decision-making to an external
engine serving as the Policy Decision Point (PDP). While our architecture supports integration with any policy engine, in this paper, we use
Cedar, which is an authorization policy language designed to be fast, safe, and analyzable~\cite{cedar2024oopsla}. It also comes with an efficient and formally verified evaluation engine. Further, a built-in policy validator can be utilized for checking the consistency of the specified policies, which is one of the main reasons behind selecting Cedar.
Such a decoupled architecture enables evolution of access control policies independent of the database system, handling policies consistently even across heterogeneous databases through a common PDP with minimal modification to the
database core.

\sysname{} retrofits ABAC on top of existing authorization models supported by the target database management system. Any requirement that can be expressed using \textit{grant} continues to be handled by DAC and one that follows the organizational structure is handled by RBAC. Additional authorizations involving attributes of the principal, resource or environment are written using ABAC as a Cedar policy. This external layer is consulted only for requests that native DAC/RBAC does not grant. Existing authorizations are therefore unaffected, and an access request can be allowed by external authorization engine only if the administrator has authored an ABAC policy permitting it. Note that externalizing authorization logic expands the Trusted Computing Base (TCB), which is a valid concern given the history of plugin vulnerabilities. However, all failure scenarios of the additional layer, including network and policy engine unavailability, leave the native decision in effect, ensuring fail-closed behavior. In a typical deployment, \sysname{} reduces the effective set of accesses, since more permissive native grants can be replaced by fine-grained attribute-based policies.

A detailed evaluation with multiple workloads shows that \sysname{} incurs only a small performance degradation for both MySQL and PostgreSQL over their native DAC/RBAC support.
With OLTP concurrency scaling, the latency overheads are
low, while maintaining
similar throughput when compared to the baseline. We also validate
that the system maintains its security guarantees under failure
conditions. \sysname{} is available as open source code with pre-built docker images. It includes
database patches, plugins, and an experimental framework for
reproducibility.

In summary, this paper makes the following
novel contributions: 
\begin{enumerate}[label=\roman*]
    \item Proposes a decoupled authorization architecture named \sysname{} for databases
    \item Introduces a direction for supporting ABAC in databases without making substantial changes to the core
    \item Develops a new MySQL plugin category and a PostgreSQL custom hook for enabling ABAC through externally specified Cedar policies
    \item Contributes to the open source community by publishing the complete implementation of ARCADE in both MySQL and PostgreSQL along with pre-built docker images
\end{enumerate}

\section{Background}
\label{sec:background}

This section provides a background on existing database authorization techniques and highlights why these are not sufficient for modern access control requirements. Subsequently, we describe ABAC, authorization decoupling using Cedar, and the database extension mechanisms we build upon.

Throughout this paper, we use a hospital schema as a motivating
example. It contains a \texttt{patient\_records} table with
columns for \texttt{patient ID}, \texttt{department} (e.g., Cardiology, Oncology),
\texttt{treating physician}, and \texttt{medical data}. Alice is a cardiologist who
should access records within her specialty, during her shift
(8:00--16:00), and that too only from the hospital internal network (10.0.0.0/8). This
policy, while simple to state, cannot be easily expressed in native database
authorization constructs.

\textbf{Native authorization in databases.} Relational database authorization primarily uses the GRANT/REVOKE model introduced for providing discretionary access~\cite{griffiths1976authorization} that allows object owners to control access through explicit
granting of rights to other users. A privilege descriptor lets a grantor associate a grantee with objects and type of access (\texttt{SELECT}, \texttt{INSERT}, \texttt{UPDATE},
\texttt{DELETE}). The \texttt{WITH GRANT OPTION} enables delegation, creating chains
of authority rooted in object owners. 

In contrast to DAC, 
an RBAC system is comprised of users, roles, and permissions, with user-to-role, permission-to-role and role-to-role mappings defined~\cite{ferraiolo1992rbac,sandhu1996role}. Instead of being granted a permission~$p$ directly, a user~$u$ has the permission if assigned to some role~$r$ that holds~$p$ either of its own or through a role-to-role mapping called a role hierarchy~\cite{sandhu2000nist}. In view of these developments in computer security research, access control in databases was also later expanded to support role-based authorizations through the use of new SQL statements like \texttt{CREATE ROLE}~\cite{chandramouliRoleBasedAccessControl1998}. Once a role is created, users can be assigned to it using the \texttt{GRANT} statement. Database access permissions like \texttt{SELECT}, \texttt{INSERT}, etc., are also assigned to a role using \texttt{GRANT}. Besides, \texttt{GRANT} can be used to assign one role to another, thus building a role hierarchy. \texttt{REVOKE} is used similarly.

Being an SQL standard, most of the current relational databases implement both DAC and RBAC. However, these models assume static, coarse-grained permissions. For instance, for the user Dr.~Alice, we can define a role named \texttt{Cardiologist} with the permission SELECT on
\texttt{patient\_records}. But this would grant access to \emph{all}
records, not just cardiology patients, and also from all locations at any point of time. Attempting to add constraints leads to
inconvenient choices. One possibility is to create
roles such as \texttt{Cardiologist\_DayShift\_OnSite}, but it causes roles to multiply combinatorially with each constraint
dimension - often called role explosion in access control literature~\cite{kuhn2010adding}. Administering and managing the same requirements using DAC is even more demanding. The alternative that the  administrators choose in practice is to retain the coarse \texttt{Cardiologist} role and leave the remaining constraints unenforced. Every cardiologist then holds \texttt{SELECT} on all records, at all times and from any location, which is more access than any of their tasks require. Such over-provisioning 
widens the damage caused by even a single compromised credential.
Another fundamental limitation is that
DAC/RBAC decisions are static, being determined at the permission or role assignment time.

ABAC addresses the above limitations by
evaluating access decisions against attributes at request time. An
ABAC policy is a relation over subject attributes (e.g., user department,
clearance), object attributes (e.g., data classification, owning
department), the requested action (e.g., Select, Insert), and environmental attributes (e.g., time,
location, risk score).
For Alice, the required policy in ABAC becomes a single expression: permit access
when \texttt{user.department = record.department} AND \texttt{time}
$\in$ [08:00, 16:00] AND \texttt{client\_ip} $\in$ 10.0.0.0/8.
Since the constraints are stated precisely, no permission is broader than the requirement it encodes. It needs to be emphasized, though, that ABAC complements DAC and RBAC rather than replacing those. Combined designs retain relatively static permissions through discretionary grants and roles, and use attribute-based rules for access to sensitive data that need stricter control~\cite{kuhn2010adding}.

An initial attempt to support ABAC in relational databases was made in ASQL (ABAC extended SQL) by incorporating statements like \texttt{CREATE RULE}, \texttt{CREATE USER ATTRIBUTE}, \texttt{REVOKE RESOURCE ATTRIBUTE}, etc~\cite{paulPosterASQLAttribute2022}. The specification was further enhanced in ASQLPLUS~\cite {meena2023efficiently} by providing support for environmental attributes like date and time of access. The authors developed a proof-of-concept implementation of ASQLPLUS in MySQL. However, unlike extending from DAC to RBAC that required minimal changes to SQL syntax and the database core, moving to ABAC in the above-mentioned approaches demands significant changes.
Furthermore, the access control logic is completely customized and incorporated into a specific version of the database, making it difficult to manage this implementation across 
evolving database core. Providing ABAC support in other databases will require customizations specific
to their respective cores. To address these challenges, our proposed architecture \sysname{} decouples authorization from the database core so that any access control model including ABAC can be retrofitted with any database management system.

\textbf{Decoupled authorization.} 
The NIST ABAC guide~\cite{hu2014guide} defines a reference architecture with four components: Policy Decision Point (PDP) evaluates requests against policies, Policy Enforcement Point (PEP) intercepts requests and enforces decisions, Policy Information Point (PIP) stores attribute values and access control rules, and Policy Administration Point (PAP) manages policies. These components need not be co-located, thereby enabling decoupled authorization, where decisions are made in one place and enforcement occurs in another.
Decoupling separates
the PEP from the PDP across a network boundary. In our context, database is the PEP,  which intercepts access requests and delegates decisions to an external PDP, if needed, via synchronous calls and finally enforces the decision. 
Such a decoupling yields three advantages. \emph{Independent evolution}: policies update directly without database restarts, and the policy engine upgrades without touching database code. \emph{Cross-database policy uniformity}: a single PDP can serve multiple, even heterogeneous, databases with unified policy administration. \emph{Minimal modification}: the architecture requires only small shim layers with native permissions staying functional.

\textbf{Cedar policy language.} Cedar~\cite{cedar2024oopsla} is a policy language
with three properties critical for database workloads. (i) Cedar
enables \emph{bounded expressivity} by supporting RBAC, ABAC, and Relationship-based Access Control (ReBAC) while restricting constructs to enable
decidable analysis. Cedar policies always terminate in contrast to
policies in general-purpose languages. The language prohibits recursion, unbounded
iteration, and user-defined functions. (ii) Cedar's Rust
implementation is highly performant as expected by modern high-throughput database systems. (iii) Cedar provides a policy validator to ensure that the ABAC policies specified are terminating and non-conflicting, which is not possible for native access control mechanisms in databases. The core semantics and implementation of Cedar itself have been formally verified using the Lean prover~\cite{cedar2024oopsla}. Listing~\ref{lst:cedar-policy} shows a Cedar policy for Alice of our example, which is simple to read but cannot be expressed in native database
authorization models. The same policy covers all other users with similar access permissions as of Alice.
\begin{prolisting}[listing options={language=Cedar}]{Cedar Policy Example}{lst:cedar-policy}
permit(
  principal,
  action == Action::"Select",
  resource == Table::"patient_records"
) when {
  principal.department == resource.department &&
  context.time >= 80000 && context.time <= 160000 &&
  context.ip.isInRange(ip("10.0.0.0/8"))
};
\end{prolisting}

Besides being a policy language, the Cedar core framework also includes an evaluation engine. In addition, Cedar agent is an HTTP server for simplified Cedar usage and deployment~\cite{permitio-cedar-agent}. It efficiently manages policy and data stores required for the Cedar evaluation engine. While ABAC policies can be specified and evaluated in Cedar, there is a need to set up a means for communicating with the database. Almost all databases provide facilities for creating extension APIs to set up such a communication path.

\textbf{Database extension mechanisms.} Databases like MySQL and PostgreSQL
provide extension mechanisms for adding functionality without modifying their core code. MySQL defines a plugin API with typed interfaces like storage engines, authentication handlers, and audit hooks~\cite{mysql8-plugins}. Plugins are shared libraries loaded at runtime, and the API has remained stable across MySQL~8.x releases. However, MySQL lacks an authorization plugin
interface, \ie, plugins can observe queries but not influence
authorization decisions.
PostgreSQL provides hooks, which are function pointers that extensions can
override. Relevant hooks include \texttt{object\_access\_hook}
(invoked on object access), \texttt{ExecutorCheckPerms\_hook} (before
query execution), and \texttt{ProcessUtility\_hook} (for DDL
statements). Extensions can modify behavior or reject
operations. However, none of the existing hooks in PostgreSQL provide an authorization intercept point.
Both the databases allow extensions to \emph{observe} authorization, but
neither provides a clean interface to \emph{extend} it. The audit plugins of MySQL cannot grant permissions, while the hooks of PostgreSQL are invoked before native checks and not after denial.

\section{ARCADE Architecture and Design}
\label{sec:design}

This section presents the design of \sysname{}. We begin with a high-level overview, then introduce
our threat model, and finally develop the architectural mechanisms that achieve the stated objectives.

\subsection{Overview}
\label{subsec:overview}
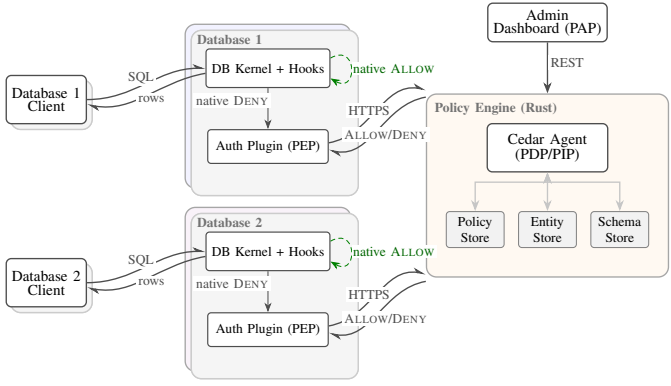
\begin{figure}[t]
  \centering
\resizebox{\linewidth}{!}{
\begin{tikzpicture}[
    >=Stealth,
    node distance=0.6cm and 0.8cm,
    comp/.style={draw=black!70, rounded corners=3pt, minimum height=0.85cm,
        minimum width=2.0cm, align=center, font=\small, fill=white, line width=0.8pt},
    smallcomp/.style={draw=black!70, rounded corners=2pt, minimum height=0.7cm,
        minimum width=2.2cm, align=center, font=\footnotesize, fill=white, line width=0.7pt},
    store/.style={draw=black!50, rounded corners=2pt, minimum height=0.6cm,
        minimum width=1.2cm, align=center, font=\footnotesize, fill=gray!10, line width=0.6pt},
    zone/.style={draw=black!35, thick, rounded corners=6pt},
    shadow/.style={draw=black!20, thick, rounded corners=6pt, fill=black!4},
    arr/.style={->, thick, >=Stealth, black!70},
    arrdbl/.style={<->, thick, >=Stealth, black!70},
    commline/.style={thick, black!70},
    reqhalf/.style={commline, -{Stealth[harpoon]}},
    resphalf/.style={commline, -{Stealth[harpoon]}},
    storelink/.style={<->, gray!45, thick, >=Stealth},
    lbl/.style={font=\footnotesize, fill=white, inner sep=1.5pt},
    zonelbl/.style={font=\footnotesize\bfseries, inner sep=1.2pt, text=black!55},
    elldots/.style={font=\large, text=black!50},
    bypass/.style={->, densely dashed, green!50!black, semithick},
    bypasstrip/.style={font=\footnotesize, text=green!40!black, fill=white, inner sep=1pt},
]

\node[comp, minimum height=1.0cm, minimum width=1.6cm]
    (client1) at (0, 1.6) {Database 1\\[-2pt]Client};

\begin{scope}[on background layer]
    \fill[shadow, fill=black!6]
        ([xshift=3pt, yshift=-3pt]client1.south west) rectangle
        ([xshift=3pt, yshift=-3pt]client1.north east);
    \draw[shadow]
        ([xshift=3pt, yshift=-3pt]client1.south west) rectangle
        ([xshift=3pt, yshift=-3pt]client1.north east);
\end{scope}

\node[smallcomp, minimum width=2.5cm, minimum height=0.80cm]
    (kernel1) at (4.6, 2.25) {DB Kernel + Hooks};
\node[smallcomp, minimum width=2.5cm, minimum height=0.70cm, below=0.85cm of kernel1]
    (plugin1) {Auth Plugin (PEP)};

\begin{scope}[on background layer]
    \node[zone, fill=blue!5,
          fit=(kernel1)(plugin1), inner xsep=0.42cm, inner ysep=0.5cm] (db1zone) {};
    \fill[shadow, fill=black!6]
        ([xshift=4pt, yshift=-4pt]db1zone.south west) rectangle
        ([xshift=4pt, yshift=-4pt]db1zone.north east);
    \draw[shadow]
        ([xshift=4pt, yshift=-4pt]db1zone.south west) rectangle
        ([xshift=4pt, yshift=-4pt]db1zone.north east);
\end{scope}
\node[zonelbl, anchor=north west]
    at ([xshift=6pt, yshift=-5pt]db1zone.north west) {Database 1};

\draw[bypass]
    ([yshift=6pt]kernel1.east) to[out=30, in=-30, looseness=4]
    node[bypasstrip, right, pos=0.5]
    {native \textsc{Allow}}
    ([yshift=-6pt]kernel1.east);

\draw[arr, semithick] (kernel1.south) --
    node[lbl, left, xshift=2pt, pos=0.3] {native \textsc{Deny}} (plugin1.north);


\node[comp, minimum height=1.0cm, minimum width=1.6cm]
    (client2) at (0, -2.2) {Database 2\\[-2pt]Client};

\begin{scope}[on background layer]
    \fill[shadow, fill=black!6]
        ([xshift=3pt, yshift=-3pt]client2.south west) rectangle
        ([xshift=3pt, yshift=-3pt]client2.north east);
    \draw[shadow]
        ([xshift=3pt, yshift=-3pt]client2.south west) rectangle
        ([xshift=3pt, yshift=-3pt]client2.north east);
\end{scope}

\node[smallcomp, minimum width=2.5cm, minimum height=0.80cm]
    (kernel2) at (4.6, -1.55) {DB Kernel + Hooks};
\node[smallcomp, minimum width=2.5cm, minimum height=0.70cm, below=0.85cm of kernel2]
    (plugin2) {Auth Plugin (PEP)};

\begin{scope}[on background layer]
    \node[zone, fill=violet!5,
          fit=(kernel2)(plugin2), inner xsep=0.42cm, inner ysep=0.5cm] (db2zone) {};
    \fill[shadow, fill=black!6]
        ([xshift=4pt, yshift=-4pt]db2zone.south west) rectangle
        ([xshift=4pt, yshift=-4pt]db2zone.north east);
    \draw[shadow]
        ([xshift=4pt, yshift=-4pt]db2zone.south west) rectangle
        ([xshift=4pt, yshift=-4pt]db2zone.north east);
\end{scope}
\node[zonelbl, anchor=north west]
    at ([xshift=6pt, yshift=-5pt]db2zone.north west) {Database 2};

\draw[bypass]
    ([yshift=6pt]kernel2.east) to[out=30, in=-30, looseness=4]
    node[bypasstrip, right, pos=0.5]
    {native \textsc{Allow}}
    ([yshift=-6pt]kernel2.east);

\draw[arr, semithick] (kernel2.south) --
    node[lbl, left, xshift=2pt, pos=0.3] {native \textsc{Deny}} (plugin2.north);

\node[comp, minimum width=2.5cm, minimum height=1cm]
    (agent) at (10.4, 0.6) {Cedar Agent\\[-1pt](PDP/PIP)};

\node[store, below=0.8cm of agent, xshift=-1.5cm] (policystore) {Policy\\Store};
\node[store, below=0.8cm of agent]                 (entitystore) {Entity\\Store};
\node[store, below=0.8cm of agent, xshift=1.5cm]   (schemastore) {Schema\\Store};

\begin{scope}[on background layer]
    \node[zone, fill=orange!5,
          fit=(agent)(policystore)(entitystore)(schemastore),
          inner xsep=0.4cm, inner ysep=0.6cm] (enginezone) {};
\end{scope}
\node[zonelbl, anchor=north west]
    at ([xshift=4pt, yshift=-4pt]enginezone.north west) {Policy Engine (Rust)};

\node[comp, minimum width=1.8cm, minimum height=0.8cm]
    (admin) at (10.4, 3.2) {Admin\\[-2pt]Dashboard (PAP)};

\draw[reqhalf] (client1.east) to[out=0, in=180]
    node[lbl, above, pos=0.45] {SQL} (kernel1.west);
\draw[resphalf] ([yshift=-3pt]kernel1.west) to[out=190, in=-10]
    node[lbl, below, pos=0.45] {rows} ([yshift=-3pt]client1.east);

\draw[reqhalf] (client2.east) to[out=0, in=180]
    node[lbl, above, pos=0.45] {SQL} (kernel2.west);
\draw[resphalf] ([yshift=-3pt]kernel2.west) to[out=190, in=-10]
    node[lbl, below, pos=0.45] {rows} ([yshift=-3pt]client2.east);

\draw[reqhalf] ([yshift=2.2pt]plugin1.east) to[out=10, in=170]
    node[lbl, above, pos=0.40] {HTTPS}
    ([yshift=2.2pt]enginezone.north west);
\draw[resphalf] ([yshift=-2.2pt]enginezone.north west) to[out=190, in=-10]
    node[lbl, below, pos=0.40, yshift=-3pt] {\textsc{Allow}/\textsc{Deny}}
    ([yshift=-2.2pt]plugin1.east);

\draw[reqhalf] ([yshift=2.2pt]plugin2.east) to[out=10, in=170]
    node[lbl, above, pos=0.40] {HTTPS}
    ([yshift=2.2pt]enginezone.south west);
\draw[resphalf] ([yshift=-2.2pt]enginezone.south west) to[out=190, in=-10]
    node[lbl, below, pos=0.40, yshift=-3pt] {\textsc{Allow}/\textsc{Deny}}
    ([yshift=-2.2pt]plugin2.east);

\draw[arr] (admin.south) -- node[lbl, right, pos=0.4] {REST} (enginezone.north);

\draw[storelink] (agent.south) -- ++(0,-0.2) -| (policystore.north);
\draw[storelink] (agent.south) -- (entitystore.north);
\draw[storelink] (agent.south) -- ++(0,-0.2) -| (schemastore.north);

\end{tikzpicture}}
\caption{\sysname{} architecture showing the external authorization path.}
\label{fig:architecture}
\end{figure}

%
\sysname{} interposes an external PDP into the
database authorization path. Unlike prior approaches that require invasive
core modifications or proxies that are bypassable, our architecture achieves ABAC
expressiveness through minimal database changes,
as illustrated in Figure~\ref{fig:architecture}.
Such a design separates enforcement (inside the database) from decision (in an
external Cedar agent), following the NIST reference
model~\cite{hu2014guide}.  Three components implement this
separation. The PEP is a database
plugin that intercepts authorization checks after native DAC/RBAC denial,
marshals requests, and translates responses. The PDP is the Cedar agent, evaluating requests against policies. The PIP, also housed in the Cedar agent, maintains the entity
store including user, database, and table names, along with their attributes. These are synchronized
from the database utilizing hooks such as object access hook in PostgreSQL and audit plugin in MySQL monitoring the Data Definition Language (DDL) statements. It needs to be emphasized that users and all objects are added and fully managed at one point only, \ie, within the database, and are seamlessly propagated to the Cedar agent using the plugins built on top of the above-mentioned interfaces. 

The authorization workflow proceeds as follows. When a client
issues a query, the database first evaluates it with native DAC/RBAC
permissions. If the native authorization permits, the query is
executed immediately without consulting the external Cedar engine. Only
when the native access control mechanisms deny the access, the external authorization hook gets invoked, and the decision returned by the Cedar agent is enforced as described in Sub-section~\ref{subsec:protocol}. Only if a fine-grained ABAC rule in Cedar is satisfied, access is granted. If no such rule exists or any error condition occurs, native authorization decision is preserved. Such a design
guarantees fail-closed behavior, \ie, failures in the external path never
grant access.
This decoupled architecture of \sysname{} yields the benefits described in
Section~\ref{sec:intro}.

\subsection{Threat Model}
\label{sec:threat-model}
In this sub-section, we present the system model, define the adversaries we defend against, and state our security objectives.

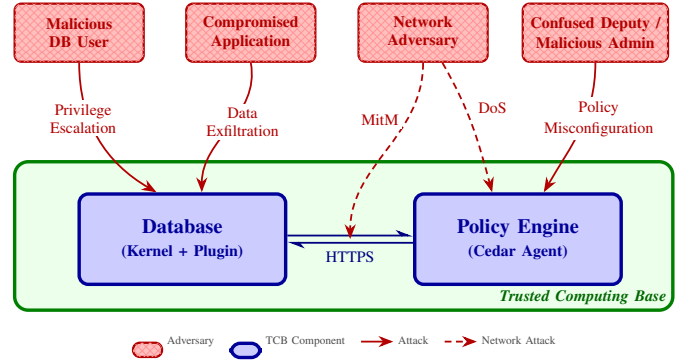
\begin{figure}[t]
  \centering
  \resizebox{\linewidth}{!}{
\begin{tikzpicture}[
    >=Stealth,
    adversary/.style={
        draw=red!70!black,
        fill=red!25,
        postaction={pattern=crosshatch, pattern color=red!40},
        rounded corners=3pt,
        line width=1.0pt,
        minimum width=2.0cm,
        minimum height=0.9cm,
        align=center,
        font=\scriptsize\bfseries,
        text=red!70!black
    },
    tcbcomp/.style={
        draw=blue!60!black,
        fill=blue!20,
        rounded corners=4pt,
        line width=1.5pt,
        minimum width=3.2cm,
        minimum height=1.4cm,
        align=center,
        font=\small\bfseries,
        text=blue!60!black
    },
    attack/.style={
        ->, thick, red!70!black, >=Stealth
    },
    attackdash/.style={
        ->, thick, red!70!black, densely dashed, >=Stealth
    },
    attacklbl/.style={font=\scriptsize, text=red!70!black, align=center, fill=white, inner sep=1.5pt},
    flow/.style={
        thick, blue!50!black
    },
    flowreq/.style={flow, -{Stealth[harpoon]}},
    flowresp/.style={flow, -{Stealth[harpoon,swap]}},
    flowlbl/.style={font=\scriptsize, text=blue!50!black},
    tcbzone/.style={draw=green!50!black, fill=green!8, rounded corners=6pt, line width=1.5pt},
]

\node[adversary] (maluser) at (1, 3.8) {Malicious\\DB User};
\node[adversary] (compapp) at (3.66, 3.8) {Compromised\\Application};
\node[adversary] (netadv) at (6.33, 3.8) {Network\\Adversary};
\node[adversary] (confdep) at (9, 3.8) {Confused Deputy /\\[1pt]{\scriptsize Malicious Admin}};

\begin{scope}[on background layer]
\node[tcbzone, fit={(0.0,-0.5)(10.2,1.8)}, inner sep=0pt] (tcb) {};
\end{scope}
\node[font=\scriptsize\bfseries\itshape, text=green!40!black, anchor=south east] at (tcb.south east) {Trusted Computing Base};

\node[tcbcomp] (database) at (2.6, 0.6) {Database\\{\scriptsize (Kernel + Plugin)}};
\node[tcbcomp] (engine) at (7.8, 0.6) {Policy Engine\\{\scriptsize (Cedar Agent)}};

\draw[flowreq] ([yshift=1.6pt]database.east) --
    ([yshift=1.6pt]engine.west);
\draw[flowreq] ([yshift=-1.6pt]engine.west) -- node[flowlbl, below, pos=0.5] {HTTPS}
    ([yshift=-1.6pt]database.east);


\draw[attack] (maluser.south) to[out=-90, in=150]
    node[attacklbl, pos=0.35, left, xshift=14pt] {Privilege\\Escalation} ([xshift=-0.4cm]database.north);

\draw[attack] (compapp.south) to[out=-80, in=90]
    node[attacklbl, right, pos=0.45, xshift=-14pt] {Data\\Exfiltration} ([xshift=0.3cm]database.north);

\coordinate (channel) at ($(database.east)!0.5!(engine.west)$);
\draw[attackdash] (netadv.south) to[out=-90, in=90]
    node[attacklbl, left, pos=0.3, xshift=-2pt] {MitM} (channel);

\draw[attackdash] ([xshift=0.3cm]netadv.south) to[out=-60, in=105]
    node[attacklbl, right, pos=0.35, xshift=4pt] {DoS} ([xshift=-0.4cm]engine.north);

\draw[attack] (confdep.south) to[out=-90, in=60]
    node[attacklbl, right, pos=0.4, xshift=-20pt] {\shortstack{Policy\\Misconfiguration}} ([xshift=0.4cm]engine.north);

\coordinate (legendanchor) at ($(tcb.south west)!0.5!(tcb.south east)$);
\node[font=\tiny, text=gray!60!black, anchor=north, yshift=-0.25cm] at (legendanchor) {
    \tikz[baseline=-0.5ex]{\node[adversary, minimum width=0.45cm, minimum height=0.22cm, font=\tiny, inner sep=1pt] {};} Adversary \quad
    \tikz[baseline=-0.5ex]{\node[tcbcomp, minimum width=0.45cm, minimum height=0.22cm, font=\tiny, inner sep=1pt] {};} TCB Component \quad
    \tikz[baseline=-0.5ex]{\draw[red!70!black, thick, ->, >=Stealth] (0,0) -- (0.5,0);} Attack \quad
    \tikz[baseline=-0.5ex]{\draw[red!70!black, thick, densely dashed, ->, >=Stealth] (0,0) -- (0.5,0);} Network Attack
};

\end{tikzpicture}}
\caption{Threat model of \sysname{} with the adversary classes and their attack vectors.}
\label{fig:threat-model}
\end{figure}

\textbf{System model.} Figure~\ref{fig:threat-model}
illustrates the trust boundaries and attack vectors. There are four entities that participate in
authorization decision making, namely (a) The database client, which is an application or
a user issuing SQL queries. It is untrusted and may be compromised or
malicious. (b) The database kernel (MySQL or PostgreSQL core, including our authorization hooks) and an in-process authorization plugin are trusted, and both lie within the TCB. The plugin is trusted to correctly translate database events to Cedar requests, communicated via HTTPS, optionally enabling mTLS.
(c) The Cedar agent is an external service, also within the TCB, that is trusted to correctly evaluate policies and return accurate decisions.
(d) The policy administrator, who would be responsible for attribute value assignment, policy creation, and auditing. Administrator accesses the Cedar agent using an Admin Dasboard (PAP in Figure~\ref{fig:architecture}), which is an untrusted application along with the administrator themselves (Malicious Admin in Figure~\ref{fig:threat-model}).

Our architecture expands the TCB beyond the native database kernel. Table~\ref{tab:tcb} enumerates each component and our justification for trusting it. As mentioned in the table, we trust the Authorization Dispatcher, which functions as a plugin registry since it expands within the database kernel. The Cedar authorization plugin, however, is enabled at the discretion of the database admin and is trusted as per the implementation and auditing done by the admin. Cedar agent is another trusted application with memory safety by design due to Rust implementation.
The communication channel between the authorization plugin and the Cedar agent is prone to attacks, and we therefore recommend enabling the mTLS support in the plugin to make it trusted. We rely on some of the formally verified properties of Cedar as mentioned in Section~\ref{sec:background}, which are default deny, explicit deny overrides, and bounded evaluation.
Cedar's skip-on-error semantics ensures that runtime errors evaluate to \textsc{Deny} rather than crashing.

\begin{table}[t]
\centering
\footnotesize
\caption{Trusted Computing Base components and trust justifications.}
\label{tab:tcb}
\rowcolors{2}{rowgray}{white}
\begin{tabular}{@{}p{2.2cm}p{2.0cm}p{3.0cm}@{}}
\toprule
\rowcolor{headergray}
\textbf{Component} & \textbf{Function} & \textbf{Trust Justification} \\
\midrule
Database Kernel & Hook invocation & Production-hardened; hooks add $\sim$300 LoC \\
Auth.\ Dispatcher & Plugin registry & Minimal logic ($\sim$500 LoC); simple semantics \\
Cedar Plugins & Request marshaling & Stateless (MySQL $\sim$1,500, PostgreSQL $\sim$1,100 LoC); no policy logic \\
Cedar Agent & Policy evaluation & Verified core; Rust \\
Network Channel & Transport & TLS with mutual authentication \\
\bottomrule
\end{tabular}
\end{table}

\textbf{Adversary.}  We consider four adversary classes, distinguished
by position and capabilities. Malicious users are legitimate users with valid credentials who attempt unauthorized access by issuing arbitrary SQL queries, exploiting application vulnerabilities, probing authorization boundaries, and measuring timing differences. Compromised applications
represent attackers who have obtained application-layer database
credentials. They can issue any query that the application could, and attempt
data ex-filtration or privilege escalation within its scope. Network adversaries are attackers on the network path
between the database and the Cedar agent, who can eavesdrop on unencrypted
traffic, perform man-in-the-middle attacks, inject malicious
responses, and cause network partitions. Confused deputies are administrators who unintentionally
define overly permissive policies, create incorrect mappings, or misconfigure schemas. None of these adversaries can modify database binaries or plugins, bypass the authorization hooks or schema validation, override a \texttt{forbid} with a \texttt{permit}, or break TLS. We exclude collusion involving compromised TCB
components. Attackers who control the database kernel or Cedar agent are also outside our threat model.

\textbf{Security objectives.} Given the adversaries, the goal of \sysname{} is to guarantee the following five properties. \emph{Non-bypass}: All authorization decisions for covered operations
must pass through the authorization dispatcher. No code path may
access protected resources without invoking registered plugins. \emph{Grant preservation}: External authorization can never revoke access granted by native DAC/RBAC. If native
authorization permits access, the query proceeds without even invoking
external authorization, so neither a compromised PDP nor a failure in the external path can lock out legitimate users. \emph{Explicit authorization}: An access request is permitted beyond DAC/RBAC only if an administrator-authored Cedar policy permits it, since Cedar denies by default and an explicit \texttt{forbid} overrides any \texttt{permit}.
\emph{Fail-closed}: System failures like network partitions, agent crashes,
timeouts, and malformed responses result in denial, not grant. Combined
with Cedar's skip-on-error semantics, the system never fails open. \emph{Scope limitation}: Authorization decisions are scoped to specific
(principal, resource, action) tuples. Further, policies for one database do not
impact another unless explicitly configured to do so.

\textbf{Assumptions and Non-Goals.}
Our security guarantees require TLS~1.3 with certificate validation for
database-to-agent communication, clock synchronization via NTP for
time-based policies, binary integrity (no rootkits or supply chain
attacks), and network isolation for the Cedar agent. We further assume that the database kernel is free of bugs and correctly invokes our hooks. \sysname{} does not defend against root attackers who can disable plugins or read process memory, against information leakage through authorization timing or access patterns, or against denial of service, for which standard rate limiting and capacity planning apply. These are examined further in Sub-section~\ref{subsec:residual-risks}.

\subsection{Layered Policy Composition}
\label{subsec:post-denial}

\sysname{} treats DAC and RBAC natively supported in databases, and Cedar as three layers of one composite authorization model. Each access requirement is expressed in the layer that can state it without granting more than the requirement needs, \ie, a \textit{grant} where a grant suffices, a role where the permission follows the organizational structure, and a Cedar policy where the decision depends on attributes of the principal, resource or environment. Least privilege is thus preserved irrespective of how context-dependent the requirement is. 
The remaining design question is when each layer is evaluated, since it is this ordering that realizes the security objectives stated in the last sub-section.
Invoking the external layer on every request, \ie, using pre-check hooks placed
before native DAC/RBAC evaluation, can create a precarious situation. The
external policy engine could then deny queries that native permissions would have
allowed. A compromised or malfunctioning agent would then be able to lock out legitimate users, violating grant preservation. We reject this approach entirely. Instead, the layers are evaluated in order. Native checks run first, and our
hooks fire only \emph{after} native checks do not give access but \emph{before}
an access privilege violation error is returned to the client. We refer to this as \emph{post-native} hook
placement. It 
guarantees grant preservation by construction, since external authorization is never invoked when native authorization succeeds. The external system cannot
interfere with natively permitted operations because it never sees them. In the other direction, Cedar denies by default, so a DAC/RBAC denial stands unless the administrator has authored a policy that permits the request, which
is the explicit authorization objective. The PEP does not have any policy
logic beyond only request marshalling and response translation, limiting its footprint.

Returning to our example, consider Alice is attempting to access oncology patient records, which is outside her specialty. If Alice has an SQL \textsc{grant} (from a DAC/RBAC policy), the query
proceeds immediately and our hooks are not invoked. If denied (as expected in this case), the post-native hook fires, constructing a Cedar authorization request with Alice's department, the current time, and her client IP. The Cedar agent evaluates relevant policies and returns its decision, perhaps finding one that permits cross-specialty access during on-call hours.

\subsection{Authorization Flow in \sysname{}}
\label{subsec:protocol}

For a query that native database authorization denies, similar to the
example given above, \sysname{} follows three critical decision points:
native authorization, external evaluation, and response translation. Each
point is designed to uphold our security objectives.

\textbf{Decision Point 1: Native Authorization.}
The database first performs its standard privilege checks. Only when there is no native authorization, does the post-native hook fire. Consequently, queries operating within native permissions incur zero overhead in the \sysname{} architecture.

\textbf{Decision Point 2: External Evaluation.}
The plugin constructs a Cedar authorization request having four
elements: the \emph{principal} (database user mapped to a Cedar entity), the \emph{action} (SELECT, INSERT, DROP, etc.), the \emph{resource} (database object with hierarchical
relationships preserved), and the \emph{context} (environmental attributes
including current time, day of week, client IP, and connection metadata).
The Cedar agent efficiently evaluates all applicable policies against this request, ensuring that such external authorization does not introduce a performance bottleneck.

\textbf{Decision Point 3: Response Translation.}
The plugin translates the Cedar agent's response into a dispatcher decision.
Cedar \textsc{Allow} maps to \textsc{Grant}, overriding the native denial.
Cedar \textsc{Deny} maps to \textsc{Deny}, explicitly preserving the
denial for logging. Errors like network failures, timeouts, malformed
responses, and agent unavailability map to \textsc{Ignore}, which also preserves the native denial. An attacker therefore gains nothing by causing network failures, since the database keeps functioning with its native access control settings.

\subsection{Entity Synchronization}
\label{subsec:entity-sync}

Cedar policies reference database objects like users, tables and database schemas by
name and attribute. For policies to evaluate correctly, the Cedar agent
must maintain an up-to-date view of the database's entity structure. It is also equally important that as and when policies or attributes change at the agent, any database-side authorization cache impacted by them must be invalidated to prevent stale decisions. This bi-directional synchronization poses a fundamental challenge. While databases push schema changes \emph{outward} to the agent, the latter pushes invalidations
\emph{inward} to databases.

\textbf{Outward Sync: Schema to Agent.}
We capture schema changes at the SQL DDL layer and propagate them to the Cedar
agent. In MySQL, a DDL audit plugin intercepts \texttt{CREATE},
\texttt{ALTER} and \texttt{DROP} statements for users, databases and
tables. In PostgreSQL, the extension registers an \texttt{object\_access\_hook} to detect schema modifications.

\textbf{Inward Sync: Invalidation to Databases.}
When policies or entities change at the Cedar agent, cached authorization
decisions in the database become stale. The agent maintains a registry of
database connections where it can push cache reset commands. MySQL uses a
global variable \textit{flush} while PostgreSQL requires a function call. In
the default configuration, the agent triggers invalidation synchronously with
policy updates. If any invalidation fails, the entire policy change rolls
back. This coupling ensures stale caches never serve decisions based on
outdated policies. Deployments prioritizing availability can instead permit
partial invalidation with bounded staleness.

\subsection{Policy Decision Point}
\label{subsec:cedar-agent}

Cedar agent~\cite{permitio-cedar-agent} is an HTTP server implemented in Rust that unifies policy decision, entity management, and administration. Built on top of Cedar, the agent inherits Cedar's formal guarantees as mentioned in Section~\ref{sec:intro}. 
We modified this implementation to provide the operational infrastructure required for integrating it with the authorization and entity synchronization plugins, while maintaining authorization cache consistency. 
Four design decisions are central to its operation.

\textbf{Schema-First validation.}
Policy and entity updates are validated against the current Cedar schema before acceptance. 
Those that invalidate the existing configurations are rejected. Consider Alice's hospital deploying a policy referencing a \texttt{specialty} attribute on physicians. If the schema does not define this attribute for \texttt{MySQL::User}, the policy is rejected at submission, thus preventing policy drift where policies reference non-existent attributes and silently fail to match.

\textbf{Policy validation.}
In Section~\ref{sec:intro} we discussed about the policy validator provided by Cedar. Cedar agent integrates this by default and ensures that the administrator cannot misconfigure policies at any point, and all the policies are terminating.

\textbf{Synchronous mutation with invalidation.}
Every policy or entity update acquires a global lock, performs the change, and triggers cache invalidation across all registered databases. In our evaluation, this lock is held for under 5\,ms on average, since policy updates are rare administrative operations, not on the hot query path. In the default configuration, if any database fails to acknowledge invalidation, the entire update rolls back. This coupling ensures every database's cache reflects the current policy set at all times. The cost is reduced availability under network partitions; deployments that prefer availability can opt into bounded-staleness invalidation.

\textbf{Memory-Safe implementation.}  Rust provides memory safety
guarantees that eliminate buffer overflows, use-after-free, and memory corruption, which are some of the most prominent vulnerabilities that have historically plagued database plugins.


\section{Implementation}
\label{sec:implementation}

We implemented \sysname{} by adding only about 2,600 lines of
database-side plugin/extension code for MySQL and PostgreSQL taken
together. This includes a MySQL authorization plugin in C++
($\sim$1,500~LoC) and a PostgreSQL extension in C ($\sim$1,100~LoC).
These figures exclude the roughly 800~LoC of kernel-side hook
instrumentation and dispatcher logic summarized in Table~\ref{tab:tcb}.
%
%
We target MySQL~8.0.43 and PostgreSQL~17. Our source codes are publicly available~\cite{archarcade-zenodo-artifact}.

\textbf{MySQL plugin interface}. The plugin interface of MySQL supports authentication, auditing and storage engines. However, no mechanism exists for external authorization decisions. Authorization logic is scattered across approximately 15 functions (Refer to Table~\ref{tab:hook-coverage}), each returning a boolean with no extensibility points.
%
We, therefore, designed a new plugin interface requiring only a few lines of core changes. The interface introduces a three-valued return type:
\textsc{Grant} (override native denial), \textsc{Deny} (preserve
denial for audit), and \textsc{Ignore} (fall back to native
behavior). Since the authorization dispatcher (see Table~\ref{tab:tcb}) is invoked only after native denial, returning
\textsc{Ignore} preserves that denial, thus ensuring any error results in denial rather than granting access.
%
We instrumented all 15 privilege-check functions with post-native
hooks.

\textbf{PostgreSQL hooks}. In contrast to MySQL, PostgreSQL exposes several extensibility
hooks. The relevant ones for our setting are
\texttt{ExecutorCheckPerms}, \texttt{object\_access} and
\texttt{ProcessUtility}. However, none of these provide an
intercept point for access control decisions (still used for the purposes mentioned in Table~\ref{tab:hook-coverage}).
We traced the authorization flow and identified a chokepoint: all privilege checks flow through \texttt{aclmask()} in \texttt{aclchk.c}, returning a bitmask of granted permissions. This centralized design enables a single hook insertion with only a few changes to the core.
The bitmask model of PostgreSQL enforces grant preservation at the machine level via bitwise OR (Listing \ref{lst:bitwise-monotonicity}). The expression \texttt{result |= mask} provides a machine-checked guarantee, \ie, bitwise OR can only set permission bits, never clear
them. Even a buggy or malicious hook cannot revoke native privileges
through this interface.

\begin{prolisting}{Bitwise Monotonicity Guarantee}{lst:bitwise-monotonicity}
result = aclmask(...);  // Native check
sufficient = (result & mask) == mask;

if (!sufficient && authorization_hook) {
  hook_result = (*authorization_hook)(&info);
  if (hook_result == PG_AUTH_RESULT_GRANT)
   result |= mask;  // Can only ADD bits
}
\end{prolisting}

The current prototype covers DML, database/schema/table/column access, routine
execution, and PostgreSQL utility/DDL paths that traverse the listed hooks.
Operations outside these privilege-check paths remain governed solely by native
DAC/RBAC. In case of MySQL, this includes administrative commands such as
\texttt{LOCK TABLES}, \texttt{FLUSH}, \texttt{SHOW} statements, and
stored-procedure compilation.
\begin{table}[t]
\centering
\footnotesize
\caption{Authorization hook locations for each database.}
\label{tab:hook-coverage}
\rowcolors{2}{rowgray}{white}
\begin{tabular}{@{}llp{3.0cm}@{}}\toprule
\rowcolor{headergray}
\textbf{Database} & \textbf{Hook Point} & \textbf{Operations} \\
\midrule
  MySQL & \texttt{check\_access} & USE, CREATE/DROP DB \\
  & \texttt{check\_grant} & SELECT, INSERT, etc. \\
  & \texttt{check\_column\_grant} & Column access \\
  & \texttt{check\_routine\_access} & Procedures, functions \\
  & \emph{($\sim$11 more functions)} & Other privilege checks \\
\midrule
  PostgreSQL & \texttt{aclmask()} & Table/view, DB, Schema, Function, Type, Column \\
  & \texttt{object\_access\_hook} & DDL entity updates \\
  & \texttt{ProcessUtility\_hook} & Utility commands \\
\bottomrule
\end{tabular}
\end{table}

\textbf{Privilege granularity mismatch.}  SQL privilege checks often
involve multiple permissions simultaneously. For instance, a statement like \texttt{INSERT INTO ... (SELECT ... FROM ...)} requires both privileges, represented as a bitmask.
Cedar evaluates one action per request with no native conjunctive
support.
We resolve this through request decomposition. For a bitmask with $n$
set bits, we issue $n$ parallel Cedar requests and then aggregate the
results. \emph{All of} requires all \textsc{Grant} responses
(matching SQL conjunctive semantics), while \emph{Any of} requires at
least one (for administrative scenarios). This enables fine-grained
policy specification, for instance, permitting SELECT on sensitive columns during business hours while permitting INSERT unconditionally.

\textbf{Cross-language communication.}  The Cedar agent runs as a
Rust process while plugins are in C/C++. We avoided Foreign Function Interface (FFI) bindings for memory safety at language boundaries, deployment complexity and network-based isolation flexibility.
We use HTTPS for communication. This introduces some latency 
but provides clean failure semantics. 
Plugins use strict
JSON serialization with SQL identifier escaping. Response parsing rejects
unexpected or missing fields as \textsc{Ignore}, with clear audit trails for
further inspection.

\textbf{Authorization caching.} External authorization adds latency
on the critical path. The MySQL plugin uses a shared sharded LRU cache
(64 shards, each guarded by its own mutex) keyed by resolved string
identifiers (user, resource, action, and time context) with configurable
TTL. Sharding reduces mutex contention under concurrent load.
The multi-process architecture of PostgreSQL requires shared memory,
so the extension uses a single \texttt{ShmemInitHash()} table guarded
by an \texttt{LWLock}.
For the PostgreSQL cache, policy changes invalidate entries by advancing
a global \texttt{last\_cache\_reset} timestamp in shared memory. Any
entry whose \texttt{created\_at} predates this timestamp is treated as a
miss on next lookup. This keeps invalidation cheap, a single write operation
rather than scanning and removing all entries under an exclusive lock.
PostgreSQL cache keys also defer OID-to-name resolution to the miss
path. Keys use raw OIDs, where OID is PostgreSQL's internal integer identifiers for
database objects, so that a cache hit requires no system-catalog access. This keeps the common case at $O(1)$ and confines catalog lookups to the slow path.

\textbf{Fail-closed error handling.}  Table~\ref{tab:failclosed}
catalogs failure modes. Every layer, namely, network, protocol, agent, and plugin, maps errors to \textsc{Ignore}. Since hooks fire only after
native denial, \textsc{Ignore} preserves that denial. No code path
converts errors to \textsc{Grant}.

\begin{table}[t]
\centering
\footnotesize
\caption{Fail-closed error handling.}
\label{tab:failclosed}
\rowcolors{2}{rowgray}{white}
\begin{tabular}{@{}lll@{}}
\toprule
\rowcolor{headergray}
\textbf{Error Class} & \textbf{Examples} & \textbf{Result} \\
\midrule
Network & Connection refused, timeout, DNS failure & \textsc{Ignore} \\
Protocol & HTTP 4xx/5xx, malformed JSON, truncated & \textsc{Ignore} \\
Agent & Crash, overload, invalid policy evaluation & \textsc{Ignore} \\
Plugin & Exception, panic, internal timeout & \textsc{Ignore} \\
\bottomrule
\end{tabular}
\end{table}

\textbf{Distributed consistency.}  If policies change, stale cached
decisions could violate access restrictions. To mitigate this, we make policy updates trigger a
coordinated sequence: validate against schema, acquire mutation lock,
apply change, 
issue cache invalidation to registered databases, wait
for acknowledgments, and finally release lock. If any database fails to
acknowledge, the policy change rolls back,
ensuring cache consistency with the authoritative policy set.
The trade-off is reduced availability during partitions. Production
deployments requiring high availability can configure partial
invalidation with bounded staleness.

\textbf{Schema synchronization.}  Outbound synchronization captures
DDL events via audit plugins (MySQL) or \texttt{object\_access\_hook}
(PostgreSQL), notifying the Cedar agent via REST. This is best-effort, 
where unreachable agents cause logged retries.
Inbound validation rejects policies referencing non-existent
entities, thus catching typos before runtime failures. Multi-database
deployments use namespace prefixes for isolation, \eg, \texttt{MySQL::hr::employees} and
\texttt{Postgres::analytics::events} remain distinct, while cross-database
policies are still possible.
Both implementations provide eventual consistency, with a sub-second window
between a DDL commit and the entity update reaching the agent. If no policy
references the new object, Cedar returns \textsc{Deny} by default. The risk
case is wildcard resource policies, \eg,
\texttt{permit(principal, action, resource in Database::"hospital")}, which
match any new table as soon as its entity propagates. Administrators relying
on such policies should apply DDL-level \texttt{REVOKE} during the
propagation window, or prefer per-object policies for sensitive tables.


\section{Evaluation}
\label{sec:evaluation}

We evaluate \sysname{} to answer three questions. (a)~What is the
performance overhead? (b)~How does \sysname{} scale under concurrent
load and policy complexity? (c)~Does the system maintain security
guarantees under failures?

\textbf{Experimental setup and methodology.}  All experiments were run on an 8-core hyper-threaded Intel
Xeon (Skylake) with 32GB RAM, Ubuntu 18.04. We deploy MySQL~8.0.43,
PostgreSQL~17, and the Cedar agent as Docker containers on the same
host. This co-located deployment for experiments is similar to our suggested production configuration and hence, the results are expected to closely align with recommended deployment scenarios. 

We use three workload types: A \emph{custom ABAC
workload} with realistic query distribution (60\% SELECT, 15\% INSERT,
15\% UPDATE, 10\% DELETE), executed on per-query basis to exercise authorization hooks; \emph{TPC-C}~\cite{sysbench-tpcc,kopytov_akopytovsysbench_2026}
via \texttt{sysbench-tpcc};  \emph{TPC-B}~\cite{Pgbench2026} via \texttt{pgbench}. We compare against
unmodified databases with semantically equivalent GRANT-based
permissions. Each configuration is run 10 times using ABBA ordering
(Alternating baseline-Cedar pairs) to account for warm-up effects and
temporal variance. We report medians with 95\% bootstrap confidence
intervals (10,000 iterations) and standard deviations. We use the
Wilcoxon signed-rank test for statistical significance ($\alpha=0.05$)
and Cliff's delta ($\delta$) for non-parametric effect size measurement. 
Experiments were run on warm systems with no other user workloads.

%
Since our initial observations revealed that the absolute latencies are higher in MySQL than PostgreSQL, we first conduct Per-query overhead, Overhead breakdown, and Concurrency scaling studies on MySQL. Next we perform TPC-C benchmarking on both MySQL and PostgreSQL. Finally, we experiment with \texttt{pgbench}, which is an industry standard TPC-B like benchmark for PostgreSQL. We report the p50, p95 and p99 latency percentiles to characterize both typical and tail behavior.
We use \texttt{toxiproxy}~\cite{ShopifyToxiproxy2026} to inject failures between the database and the Cedar agent for studying the fail-closed behavior.

\begin{table}[t]
\centering
\scriptsize
\caption{Per-operation latency overhead for MySQL (10 runs; median [95\% CI]).}
\label{tab:operation-overhead}
\rowcolors{2}{rowgray}{white}
\begin{tabular}{@{}l
            l
            l
            S[table-format=2.1, table-space-text-post=\%]@{}}
\toprule
\rowcolor{headergray}
{Op} & {Baseline (ms)} & {Cedar (ms)} & {Overhead} \\
\midrule
SELECT & 1.38 [1.36--1.42] & 1.44 [1.42--1.46] & +4.5\% \\
INSERT & 1.25 [1.23--1.28] & 1.28 [1.26--1.32] & +2.8\% \\
UPDATE & 5.24 [5.20--5.26] & 5.24 [5.20--5.28] & {$\sim$0\%} \\
DELETE & 1.33 [1.31--1.39] & 1.40 [1.39--1.43] & +5.5\% \\
JOIN & 1.45 [1.44--1.50] & 1.51 [1.49--1.53] & +3.9\% \\
\bottomrule
\end{tabular}
\end{table}

\textbf{Per-query overhead.}  We measure per-operation latency across
10 paired runs using ABBA ordering to characterize authorization cost.
Table~\ref{tab:operation-overhead} presents median latency with 95\%
bootstrap confidence intervals for each SQL operation type, comparing
baseline MySQL against Cedar-enabled MySQL. 
For no operation type is the overhead statistically significant at $\alpha$=0.05 (Wilcoxon signed-rank, Bonferroni-corrected). For high-latency operations like UPDATE, it is negligible and within measurement variance.
%
%
Fast operations show higher relative overhead (2.8--5.5\%), reflecting
that authorization (at $\sim$0.06ms) becomes more visible when base
latency is low. Across all operations, overhead ranges from $\sim$0\%
to +5.5\%, with overlapping confidence intervals indicating that the differences may be attributable to normal system variance.

\textbf{Overhead breakdown.} To pinpoint where time is spent in
authorization logic, we use the \textit{performance schema} of MySQL to examine
internal stage timings. Table~\ref{tab:overhead-breakdown} shows each
stage's share of the total per-query overhead across 10 runs.
\begin{table}[t]
\centering
\footnotesize
\caption{MySQL stage overhead breakdown (Performance Schema, 10 runs).}
\label{tab:overhead-breakdown}
\rowcolors{2}{rowgray}{white}
\begin{tabular}{@{}l
            S[table-format=2.1, table-space-text-post=\%]@{}}
\toprule
\rowcolor{headergray}
{MySQL Stage} & {Overhead} \\
\midrule
\texttt{checking permissions} & +60.7\% \\
\texttt{starting} & +12.4\% \\
\texttt{waiting for handler commit} & +11.9\% \\
\texttt{opening tables} & +9.1\% \\
\texttt{freeing items} & +3.5\% \\
\texttt{others} & +2.4\% \\
\bottomrule
\end{tabular}
\end{table}
The primary overhead appears in \texttt{checking permissions}, consistent with our design in which the authorization
plugin adds work only on the slow path where native DAC/RBAC
has already denied access. The remainder is attributable to the
\texttt{starting} stage, which covers construction of the
authorization request, and \texttt{waiting for handler commit},
which reflects round-trip latency to the Cedar agent. The
net overhead of $\sim$0.06ms per query agrees with the per-operation overhead observed in
Table~\ref{tab:operation-overhead}.

\begin{figure}[t]
\centering
\begin{tikzpicture}

\definecolor{baselinecolor}{RGB}{31, 119, 180}
\definecolor{cedarcolor}{RGB}{214, 39, 40}

\begin{axis}[
  width=\columnwidth,
  height=0.55\columnwidth,
  ybar,
  bar width=18pt,
  ymin=0,
  ymax=93000,
  ymajorgrids=true,
  grid style={gray!25, thin},
  ylabel={Throughput},
  ylabel style={font=\small},
  symbolic x coords={PostgreSQL,MySQL},
  xtick=data,
  xticklabels={PostgreSQL\\{\scriptsize(TPC-C)}, MySQL\\{\scriptsize(TPC-C)}},
  xticklabel style={align=center, font=\small},
  yticklabel style={font=\small},
  enlarge x limits=0.4,
  legend style={
    at={(0.5,1.02)},
    anchor=south,
    draw=gray!50,
    fill=white,
    fill opacity=0.95,
    font=\footnotesize,
    legend columns=2,
    column sep=8pt,
  },
  tick label style={font=\small},
]

\addplot+[
  fill=baselinecolor!70,
  draw=baselinecolor,
  very thick,
  error bars/.cd,
  y dir=both,
  y explicit,
] coordinates {
  (PostgreSQL, 76694) +- (0, 2876)
  (MySQL, 10011) +- (0, 1421)
};
\addlegendentry{Baseline}

\addplot+[
  fill=cedarcolor!70,
  draw=cedarcolor,
  very thick,
  postaction={pattern=north east lines, pattern color=cedarcolor!90},
  error bars/.cd,
  y dir=both,
  y explicit,
] coordinates {
  (PostgreSQL, 75709) +- (0, 4114)
  (MySQL, 9634) +- (0, 1501)
};
\addlegendentry{With Cedar}

\node[anchor=south, font=\scriptsize\bfseries, text=red!70!black] 
  at (axis cs:PostgreSQL, 82000) {+1.3\%};
\node[anchor=south, font=\scriptsize\bfseries, text=red!70!black] 
  at (axis cs:MySQL, 13500) {+3.9\%};

\end{axis}
\end{tikzpicture}
\caption{TPC-C performance (10 warehouses, 16 threads, 10 runs; error bars are $\pm$1 std.\ dev.).}
\label{fig:cross-database}
\end{figure}
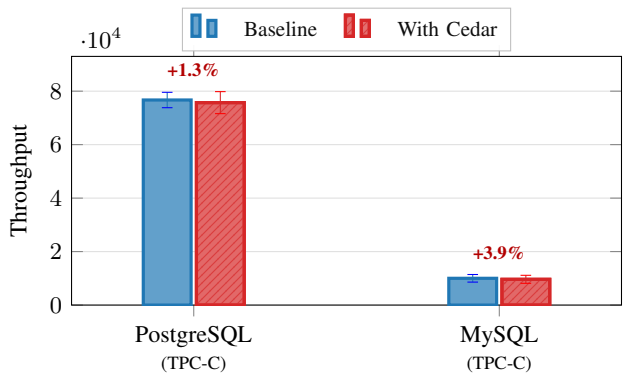

\textbf{Concurrency scaling.} A key concern for production deployments
is whether Cedar introduces bottlenecks under concurrent load. In
Figure~\ref{fig:concurrency}, we measure throughput and latency using
the \texttt{sysbench} OLTP workload with 1--16 threads.
\begin{figure}[t]
\centering
\begin{tikzpicture}
\begin{axis}[
  width=0.95\columnwidth,
  height=0.58\columnwidth,
  ybar,
  bar width=8pt,
  ymin=0,
  ymax=30000,
  ymajorgrids=true,
  grid style={gray!20},
  xlabel={Concurrent Threads},
  ylabel={Queries per Second},
  ytick={5000,10000,15000,20000,25000},
  scaled y ticks = true,
  yticklabel style={/pgf/number format/fixed},
  symbolic x coords={1,2,4,8,12,16},
  every y tick scale label/.style={
    at={(-0.14,1)}, 
    anchor=south west,
    font=\bfseries\large, 
    fill=white,
    fill opacity=0.8,
    text opacity=1,
    xshift=0.5em,
    yshift=0.5em
  },
  xtick=data,
  legend columns=2,
  legend style={
    at={(0.53,1.05)},
    anchor=south,
    draw=none,
    fill=white,
    fill opacity=0.9,
    font=\footnotesize,
    column sep=5pt
  },
  tick align=outside,
  enlarge x limits=0.12,
  clip=false,
]

\addplot+[
  fill=NavyBlue!70,
  draw=NavyBlue!90,
  mark=none,
  error bars/.cd,
  y dir=both,
  y explicit,
] table[
  col sep=comma,
  x=threads_str,
  y=baseline_qps,
  y error=baseline_qps_std,
] {figures/data/concurrency_comparison_str.csv};

\addplot+[
  fill=BrickRed!70,
  draw=BrickRed!90,
  mark=none,
  postaction={pattern=north east lines, pattern color=BrickRed!90},
  error bars/.cd,
  y dir=both,
  y explicit,
] table[
  col sep=comma,
  x=threads_str,
  y=cedar_qps,
  y error=cedar_qps_std,
] {figures/data/concurrency_comparison_str.csv};

\legend{Baseline MySQL,Cedar-Enabled MySQL}
\end{axis}
\end{tikzpicture}
\caption{Throughput scaling with concurrency (10 runs; error bars are $\pm$1 std.\ dev.).}
\label{fig:concurrency}
\end{figure}
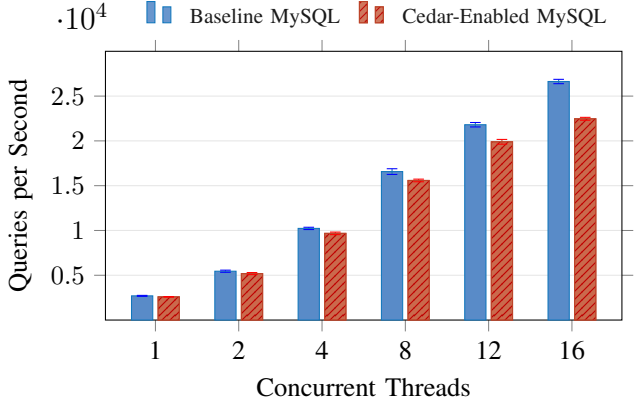
\begin{figure}[t]
\centering
%
\begin{tikzpicture}
\begin{axis}[
  width=0.95\columnwidth,
  height=0.62\columnwidth,
  ymin=0,
  ymax=25,
  ymajorgrids=true,
  grid style={gray!20},
  xlabel={Concurrent Threads},
  ylabel={Latency (ms)},
  symbolic x coords={1,2,4,8,12,16},
  xtick=data,
  legend columns=2,
  legend style={
    at={(0.5,1.05)},
    anchor=south,
    draw=none,
    fill=white,
    fill opacity=0.9,
    font=\footnotesize,
    column sep=6pt,
    /tikz/every even column/.append style={column sep=12pt},
  },
  tick align=outside,
]

\addplot+[NavyBlue,thick,solid,mark=*,mark size=2pt] table[
  col sep=comma,
  x=threads_str,
  y=baseline_p50_ms
] {figures/data/concurrency_comparison_str.csv};

\addplot+[BrickRed,only marks,mark=o,mark size=3.5pt,
  mark options={solid,line width=1pt}
] table[
  col sep=comma,
  x=threads_str,
  y=cedar_p50_ms
] {figures/data/concurrency_comparison_str.csv};

\addplot+[
  NavyBlue,thick,densely dotted,
  mark=square*,mark size=2pt,
  mark options={fill=NavyBlue!90}
] table[
  col sep=comma,
  x=threads_str,
  y=baseline_p95_ms
] {figures/data/concurrency_comparison_str.csv};

\addplot+[BrickRed,only marks,mark=square,mark size=3.5pt,
  mark options={solid,line width=1pt}
] table[
  col sep=comma,
  x=threads_str,
  y=cedar_p95_ms
] {figures/data/concurrency_comparison_str.csv};

\legend{Baseline p50, Cedar p50, Baseline p95, Cedar p95}
\end{axis}
\end{tikzpicture}
\caption{Latency percentiles vs.\ concurrency (1--16 threads).}
\label{fig:concurrency-latency}
\end{figure}
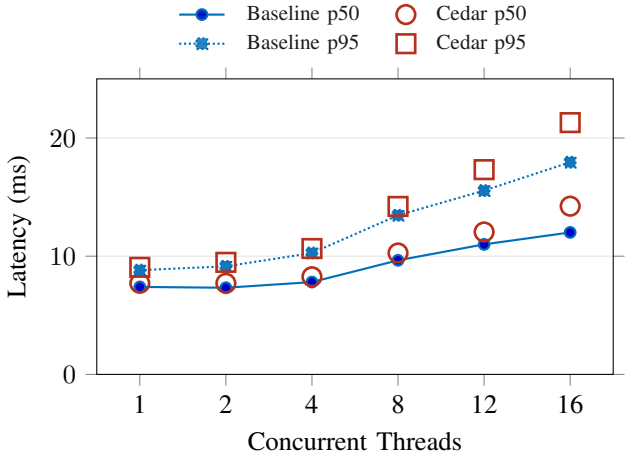
The figure shows baseline MySQL reaching 26,629
$\pm$243 Queries per Second (QPS) at 16 threads versus 22,467 $\pm$162 QPS with Cedar
(15.6\% throughput reduction). The overhead increases with concurrency:
4.0\% at 1 thread to 15.6\% at 16 threads, reflecting increased contention
for authorization resources under load.
Figure~\ref{fig:concurrency-latency} shows the query latencies plotted against the number of concurrent threads (clients) connecting to the database.
It is seen that p50 latency increases from
7.4ms to 14.2ms as threads scale from 1 to 16, with Cedar adding
2--3ms overhead at high concurrency. At p95, the additional tail latency is only 0.2--3.4ms over the same range. This \texttt{sysbench} OLTP workload is a micro-benchmark that
stresses the authorization slow path more aggressively than TPC-C. We
therefore treat Figure~\ref{fig:cross-database}, where MySQL overhead
stays below 3.9\%, as the primary end-to-end throughput reference.

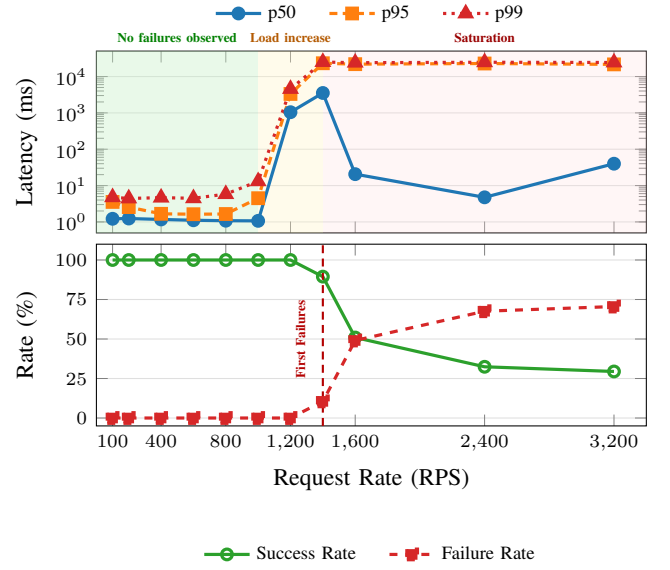
\begin{figure}[t]
\centering
\begin{tikzpicture}

\definecolor{safezone}{RGB}{200, 240, 200}
\definecolor{warnzone}{RGB}{255, 245, 200}
\definecolor{dangerzone}{RGB}{255, 220, 220}
\definecolor{p50color}{RGB}{31, 119, 180}
\definecolor{p95color}{RGB}{255, 127, 14}
\definecolor{p99color}{RGB}{214, 39, 40}
\definecolor{successcolor}{RGB}{44, 160, 44}
\definecolor{failurecolor}{RGB}{214, 39, 40}

\begin{axis}[
  name=latencyplot,
  width=\columnwidth,
  height=0.45\columnwidth,
  ylabel={Latency (ms)},
  ylabel style={font=\small},
  xmin=0, xmax=3400,
  ymin=0.5, ymax=50000,
  ymode=log,
  xtick={100,400,800,1200,1600,2400,3200},
  xticklabels={},
  ytick={1,10,100,1000,10000},
  yticklabel style={font=\scriptsize},
  ymajorgrids=true,
  grid style={gray!25, thin},
  legend style={
    at={(0.5,1.1)},
    anchor=south,
    draw=none,
    fill=none,
    font=\scriptsize,
    row sep=1pt,
    /tikz/every even column/.append style={column sep=10pt},
  },
  legend columns=-1,
  tick label style={font=\scriptsize},
  clip=false,
]

\fill[safezone, opacity=0.4] (axis cs:0,0.5) rectangle (axis cs:1000,50000);

\fill[warnzone, opacity=0.35] (axis cs:1000,0.5) rectangle (axis cs:1400,50000);

\fill[dangerzone, opacity=0.25] (axis cs:1400,0.5) rectangle (axis cs:3400,50000);

\addplot[p50color, very thick, mark=*, mark size=2pt, mark options={solid}] coordinates {
  (100, 1.22)
  (200, 1.24)
  (400, 1.17)
  (600, 1.11)
  (800, 1.08)
  (1000, 1.07)
  (1200, 1044.29)
  (1400, 3530.02)
  (1600, 20.62)
  (2400, 4.75)
  (3200, 39.96)
};
\addlegendentry{p50}

\addplot[p95color, very thick, densely dashed, mark=square*, mark size=2pt, mark options={solid}] coordinates {
  (100, 3.55)
  (200, 2.56)
  (400, 1.67)
  (600, 1.63)
  (800, 1.65)
  (1000, 4.49)
  (1200, 3319.70)
  (1400, 23212.90)
  (1600, 21924.01)
  (2400, 22841.17)
  (3200, 21608.64)
};
\addlegendentry{p95}

\addplot[p99color, very thick, dotted, mark=triangle*, mark size=2.5pt, mark options={solid}] coordinates {
  (100, 4.77)
  (200, 4.42)
  (400, 4.65)
  (600, 4.47)
  (800, 5.81)
  (1000, 13.11)
  (1200, 4550.51)
  (1400, 25044.12)
  (1600, 24060.41)
  (2400, 24723.82)
  (3200, 24371.41)
};
\addlegendentry{p99}

\node[anchor=south, font=\tiny\bfseries, text=green!50!black] 
            at (axis cs:500,50000) {No failures observed};
\node[anchor=south, font=\tiny\bfseries, text=orange!70!black] 
  at (axis cs:1200,50000) {Load increase};
\node[anchor=south, font=\tiny\bfseries, text=red!60!black] 
  at (axis cs:2400,50000) {Saturation};

\end{axis}

\begin{axis}[
  name=rateplot,
  at={(latencyplot.south west)},
  anchor=north west,
  yshift=-0.15cm,
  width=\columnwidth,
  height=0.45\columnwidth,
  xlabel={Request Rate (RPS)},
  ylabel={Rate (\%)},
  xlabel style={font=\small},
  ylabel style={font=\small},
  xmin=0, xmax=3400,
  ymin=-5, ymax=110,
  xtick={100,400,800,1200,1600,2400,3200},
  xticklabel style={font=\scriptsize},
  ytick={0,25,50,75,100},
  yticklabel style={font=\scriptsize},
  ymajorgrids=true,
  grid style={gray!25, thin},
  legend style={
    at={(0.5,-0.6)},
    anchor=north,
    draw=none,
    fill=none,
    font=\scriptsize,
    /tikz/every even column/.append style={column sep=10pt},
  },
  legend columns=-1,
  tick label style={font=\scriptsize},
  clip=false,
]

\addplot[successcolor, very thick, mark=o, mark size=2pt] coordinates {
  (100, 100)
  (200, 100)
  (400, 100)
  (600, 100)
  (800, 100)
  (1000, 100)
  (1200, 100)
  (1400, 89.47)
  (1600, 50.94)
  (2400, 32.42)
  (3200, 29.45)
};
\addlegendentry{Success Rate}

\addplot[failurecolor, very thick, dashed, mark=square*, mark size=2pt] coordinates {
  (100, 0)
  (200, 0)
  (400, 0)
  (600, 0)
  (800, 0)
  (1000, 0)
  (1200, 0)
  (1400, 10.53)
  (1600, 49.06)
  (2400, 67.58)
  (3200, 70.55)
};
\addlegendentry{Failure Rate}

\draw[red!70!black, densely dashed, thick] (axis cs:1400,-5) -- (axis cs:1400,110);
\node[anchor=south, rotate=90, font=\tiny\bfseries, text=red!70!black] 
  at (axis cs:1350,50) {First Failures};

\end{axis}

\end{tikzpicture}
\caption{Cedar agent stress test: latency percentiles (top) and success/failure rates (bottom).}
\label{fig:agent-stress}
\end{figure}

\textbf{Sensitivity to load on the Cedar agent.} The Cedar agent is
a shared resource serving all authorization requests. To characterize
its capacity limits, we stress-test the agent at 100--3,200 requests
per second (RPS), as shown in Figure~\ref{fig:agent-stress}. The agent handles 1,000 RPS with sub-15ms p99
latency. At 1,200 RPS, queuing causes p50 to spike to 1,044ms with
0\% failures. Beyond 1,400 RPS, connection exhaustion causes 10.5\%
failures, and p50 latency drops sharply because failed requests complete in milliseconds instead of waiting in queue. Successful requests at that point still experience extreme latency (p95/p99 at $\sim$22--25s), which confirms the fail-fast behavior under overload. High-throughput environments can scale horizontally by
deploying multiple agents behind a load balancer.

\begin{table}[t]
\centering
\footnotesize
\caption{Impact of injected network delay in MySQL.}
\label{tab:delay}
\rowcolors{2}{rowgray}{white}
\begin{tabular}{@{}
    S[table-format=3.0]
    S[table-format=3.2]
    S[table-format=1.2]@{}}
\toprule
\rowcolor{headergray}
{Injected Delay (ms)} & {Query Time (ms)} & {Resolution Time (ms)} \\
\midrule
0 & 5.19 & 5.19 \\
5 & 9.81 & 4.81 \\
10 & 16.09 & 6.09 \\
50 & 56.22 & 6.22 \\
100 & 106.24 & 6.24 \\
500 & 507.58 & 7.58 \\
\bottomrule
\end{tabular}
\end{table}

\textbf{Sensitivity to network delay.} Distributed deployments may
place the Cedar agent on a separate host, introducing network
latency. As illustrated in Table~\ref{tab:delay}, we inject delays via \texttt{toxiproxy} (0--500ms) to characterize
behavior. The intrinsic authorization cost is constant and predictable: subtracting the injected delay from total query latency yields a near constant 5--7ms, regardless of network conditions, enabling straightforward capacity planning.

\begin{figure}[t]
\centering
%
\begin{tikzpicture}

\begin{axis}[
  width=\columnwidth,
  height=0.60\columnwidth,
  xmin=0.5,
  xmax=5.5,
  xtick={1,2,3,4,5},
  xticklabels={$10^0$,$10^1$,$10^2$,$10^3$,$10^4$},
  xlabel={Policy Count},
  ylabel={Authorization Latency (ms)},
  ymin=0,
  ymax=2.8,
  ymajorgrids=true,
  grid style={gray!25,thin},
  tick align=outside,
  boxplot/draw direction=y,
  legend columns=2,
  legend style={
    at={(0.5,1.05)},
    anchor=south,
    draw=none,
    fill=white,
    fill opacity=0.90,
    font=\footnotesize,
    column sep=6pt
  },
]

\pgfmathsetmacro{\dx}{0.15}

\IfFileExists{figures/data/policy_scaling_boxplot_stats.csv}{}{%
  \PackageError{policy_scaling_boxplot}{Missing figures/data/policy_scaling_boxplot_stats.csv}{%
    Expected figures/data/policy_scaling_boxplot_stats.csv relative to usenix/paper/.%
  }%
}
\pgfplotstableread[col sep=comma]{figures/data/policy_scaling_boxplot_stats.csv}\policystats
\pgfplotstablegetrowsof{\policystats}
\pgfmathtruncatemacro{\lastrow}{\pgfplotsretval-1}
\ifnum\lastrow<0
  \PackageError{policy_scaling_boxplot}{Empty policy_scaling_boxplot_stats.csv}{CSV has no data rows.}%
\fi


\newcommand{\ifeqstr}[3]{%
  \edef\psbifeqstra{#1}%
  \edef\psbifeqstrb{#2}%
  \ifdefined\pdfstrcmp
    \ifnum\pdfstrcmp{\psbifeqstra}{\psbifeqstrb}=0 #3\fi
  \else\ifdefined\strcmp
    \ifnum\strcmp{\psbifeqstra}{\psbifeqstrb}=0 #3\fi
  \else
    \ifx\psbifeqstra\psbifeqstrb #3\fi
  \fi\fi
}

\pgfplotsforeachungrouped \r in {0,...,\lastrow} {
  \pgfplotstablegetelem{\r}{policy_count}\of{\policystats}\edef\pc{\pgfplotsretval}
  \pgfplotstablegetelem{\r}{system}\of{\policystats}\edef\sys{\pgfplotsretval}

  \pgfplotstablegetelem{\r}{whisker_low_ms}\of{\policystats}\edef\wl{\pgfplotsretval}
  \pgfplotstablegetelem{\r}{q1_ms}\of{\policystats}\edef\lq{\pgfplotsretval}
  \pgfplotstablegetelem{\r}{median_ms}\of{\policystats}\edef\med{\pgfplotsretval}
  \pgfplotstablegetelem{\r}{q3_ms}\of{\policystats}\edef\uq{\pgfplotsretval}
  \pgfplotstablegetelem{\r}{whisker_high_ms}\of{\policystats}\edef\wh{\pgfplotsretval}

  \pgfmathsetmacro{\xidx}{log10(\pc)+1}

  \def\drawcol{NavyBlue}
  \def\fillcol{NavyBlue!18}
  \def\usepattern{0}
  \pgfmathsetmacro{\pos}{\xidx - \dx}

  \ifeqstr{\sys}{cedar}{%
    \def\drawcol{BrickRed}
    \def\fillcol{BrickRed!18}
    \def\usepattern{1}
    \pgfmathsetmacro{\pos}{\xidx + \dx}
  }

  \edef\psbdrawcol{\drawcol}
  \edef\psbfillcol{\fillcol}
  \edef\psbusepattern{\usepattern}
  \edef\psbpos{\pos}
  \edef\psbwl{\wl}
  \edef\psblq{\lq}
  \edef\psbmed{\med}
  \edef\psbuq{\uq}
  \edef\psbwh{\wh}

  \ifnum\psbusepattern=1
    \edef\psbaddplotcmd{\noexpand\addplot+[%
      forget plot,
      boxplot prepared={
        draw position={\psbpos},
        box extend=0.25,
        lower whisker=\psbwl,
        lower quartile=\psblq,
        median=\psbmed,
        upper quartile=\psbuq,
        upper whisker=\psbwh
      },
      very thick,
      draw=\psbdrawcol,
      preaction={fill=\psbfillcol},
      pattern=north east lines,
      pattern color=BrickRed!70,
      solid,
    ] coordinates {};}
  \else
    \edef\psbaddplotcmd{\noexpand\addplot+[%
      forget plot,
      boxplot prepared={
        draw position={\psbpos},
        box extend=0.25,
        lower whisker=\psbwl,
        lower quartile=\psblq,
        median=\psbmed,
        upper quartile=\psbuq,
        upper whisker=\psbwh
      },
      very thick,
      draw=\psbdrawcol,
      fill=\psbfillcol,
      solid,
    ] coordinates {};}
  \fi
  \psbaddplotcmd
}

\addlegendimage{area legend, draw=NavyBlue, fill=NavyBlue!18, very thick}
\addlegendentry{Baseline}
\addlegendimage{area legend, draw=BrickRed, preaction={fill=BrickRed!18}, pattern=north east lines, pattern color=BrickRed!70, very thick}
\addlegendentry{With Cedar}

\end{axis}
\end{tikzpicture}
\caption{Authorization latency vs.\ policy count.}
\label{fig:policy-scaling}
\end{figure}
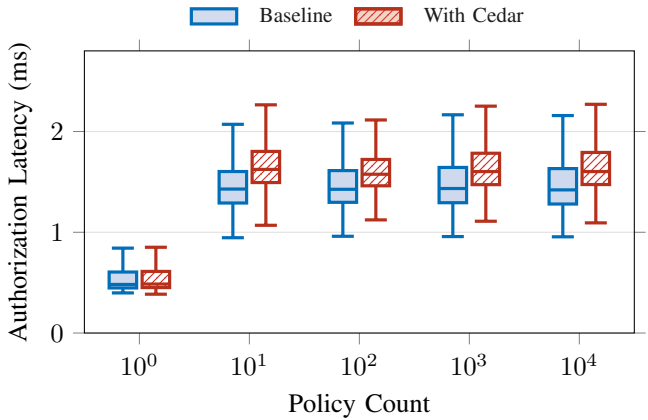

\textbf{Scaling to large policy sets.} Production deployments may
accumulate hundreds or thousands of policies. We evaluate
authorization latency with 1--10,000 policies to characterize scaling
behavior. Figure~\ref{fig:policy-scaling} shows an initial latency
jump of 3$\times$ between 1 policy ($\sim$0.48ms) and 10+ policies
($\sim$1.6ms), attributable to Cedar's policy indexing overhead.
Beyond 10 policies, latency is essentially flat, going from 1.62ms at 10
policies to 1.60ms at 10,000 policies. This is consistent with Cedar's
type-based indexing that achieves sub-linear lookup times. Tail latencies (p95/p99) show greater variance, which can be attributed to garbage collection in the Rust runtime.

\begin{table}[t]
\centering
\footnotesize
\caption{PostgreSQL TPC-B with pgbench (scale=20, 4 clients, 10 runs).}
\label{tab:pgbench}
\rowcolors{2}{rowgray}{white}
\begin{tabular}{@{}l
            l
            l
            r@{}}
\toprule
\rowcolor{headergray}
{Metric} & {Baseline} & {Cedar} & {Overhead} \\
\midrule
TPS & $7737 \pm 83$ & $7457 \pm 129$ & +3.6\% \\
Latency (ms) & $0.517$ & $0.537$ & +3.9\% \\
\bottomrule
\end{tabular}
\end{table}

\textbf{Performance with different databases.} To validate that our
approach generalizes across database engines, we evaluate TPC-C
performance on both PostgreSQL and
MySQL. Figure~\ref{fig:cross-database} shows results across 10 runs
each. PostgreSQL achieves $76{,}694 \pm 2{,}876$ Transactions per Minute (TPM) (median $\pm$
std.\ dev.) baseline versus $75{,}709 \pm 4{,}114$ TPM with Cedar,
yielding 1.3\% median throughput reduction with overlapping variance
ranges.  MySQL achieves $10{,}011 \pm 1{,}421$ TPM baseline versus
$9{,}634 \pm 1{,}501$ TPM with Cedar (3.9\% median reduction). MySQL
median latency changes only from 12.5ms to 12.7ms.
Both databases thus show overhead near measurement variance for complex transaction workloads. The slightly higher variance
in Cedar measurements reflects warm-up effects in early runs, with runs 3--10 representing steady-state performance. For TPC-B
(Table~\ref{tab:pgbench}), the overhead is 3.6\% at $7{,}457 \pm 129$ TPS, showing that
\sysname{} scales well for simple transaction workloads as well.


\section{Security Analysis}
\label{sec:security}
As noted earlier, decoupling authorization from the database kernel expands the
TCB and introduces a network boundary on
the authorization critical path. This section analyzes the security implications of these changes, building on the threat model of Sub-section~\ref{sec:threat-model} and the enforcement mechanism of Sub-section~\ref{subsec:post-denial}.

\subsection{Attack surface analysis}
Decoupling as done in \sysname{} introduces three security-relevant interfaces beyond the native database authorization path.

\textbf{In-process enforcement boundary (SQL $\rightarrow$ authorization hook).}
The primary concern at this boundary is coverage: a privileged operation could execute along a code path that does not invoke the post-native hook, violating the non-bypass objective (Sub-section~\ref{sec:threat-model}). \sysname{} addresses this by instrumenting all relevant privilege-check call sites in MySQL and by inserting a single chokepoint at \texttt{aclmask()} in PostgreSQL (Table~\ref{tab:hook-coverage}), backed by regression tests that exercise every covered operation class. The current prototype covers DML, database/schema/table/column checks, routine execution, and PostgreSQL utility/DDL paths that traverse these hooks; server-administrative operations outside those paths continue to rely on native DAC/RBAC. The residual risk is \emph{evolution}, \ie, future upstream releases may introduce new privilege-check paths that bypass existing hooks. Continuous testing against new database versions is therefore necessary.

\begin{table}[t]
\centering
\footnotesize
\caption{Network interface threats and mitigations.}
\label{tab:network-threats}
\rowcolors{2}{rowgray}{white}
\begin{tabular}{@{}p{2.0cm}p{2.3cm}p{2.3cm}@{}}
\toprule
\rowcolor{headergray}
\textbf{Threat} & \textbf{Mitigation} & \textbf{Residual Risk} \\
\midrule
Eavesdropping & TLS 1.3 & Endpoint breach \\
MITM/Tampering & Cert validation, mTLS & CA compromise \\
Replay attacks & Request timestamps ($\pm$5\,s window) & Clock skew $>$5\,s \\
Denial of service & Timeout + fail-closed & Availability \\
\bottomrule
\end{tabular}
\end{table}

\textbf{Network boundary (plugin $\leftrightarrow$ Cedar agent).}
This communication boundary is an externally reachable surface. Table~\ref{tab:network-threats} enumerates the threats it is exposed to, along with the mitigation adopted in \sysname{} and the risk that remains after the mitigation. The critical security consequence is that network-layer attacks can at worst induce availability loss and not unauthorized access, because every external-path anomaly maps to \textsc{Ignore} at the plugin boundary, preserving the native denial as mentioned in Table~\ref{tab:failclosed}.

\textbf{Policy administration boundary} 
Decoupling elevates policy administration as a prominent attack surface: an administrator can author syntactically valid policies that are semantically too permissive (\textit{confused deputy} in Figure~\ref{fig:threat-model}). \sysname{} prevents such errors through schema-first validation (Section~\ref{sec:design}) and Cedar's policy validator (Section~\ref{sec:background}), which reject undefined attribute references, type mismatches, and non-terminating constructs before deployment (Sub-section~\ref{subsec:cedar-agent}). Semantic mistakes, \ie, valid policy, and wrong intent remain possible and require operational controls (review, staged rollout, audit logging) rather than mechanism-level fixes.

\begin{table}[t]
\centering
\footnotesize
\caption{Plugin vulnerabilities and our mitigations.}
\label{tab:plugin-cves}
\rowcolors{2}{rowgray}{white}
\begin{tabular}{@{}lp{1.8cm}p{3.2cm}@{}}
\toprule
\rowcolor{headergray}
\textbf{CVE} & \textbf{Class} & \textbf{Mitigation} \\
\midrule
CVE-2020-14878~\cite{cve-2020-14878} & Memory safety & Minimal C++; Rust \\
CVE-2021-27928~\cite{cve-2021-27928} & Library loading & No dynamic loading \\
CVE-2023-21980~\cite{cve-2023-21980} & Path traversal & No file access \\
CVE-2022-2625~\cite{cve-2022-2625} & Privilege esc. & No elevated ops \\
CVE-2024-10979~\cite{cve-2024-10979} & Environment RCE & No env processing \\
\bottomrule
\end{tabular}
\end{table}

Database plugins have historically been a source of critical
vulnerabilities 
as enumerated in Table~\ref{tab:plugin-cves}. Our design mitigates
these by using minimal C++ with logic in Rust.

\subsection{Security properties}
\sysname{}'s guarantees follow from the layered composition of Sub-section~\ref{subsec:post-denial} and the fail-closed implementation of Section~\ref{sec:implementation}, which maps every external failure, \ie, timeouts, malformed responses, agent crashes, and network partitions, to \textsc{Ignore} at the plugin boundary (Table~\ref{tab:failclosed}). We now consider what they imply under adversarial conditions.
A fully compromised agent
can never revoke access that native DAC/RBAC already permits. The worst-case impact of Cedar compromise is over-permissioning within the externally governed portion of the permission space (a confidentiality risk), not lockout. An attacker who controls Cedar cannot deny access to users the database itself already permits.
When multiple Cedar agent replicas serve the same fleet, requests routed to a stale replica during a policy update may receive outdated decisions. This window is bounded by network round-trip time ($\leq$10\,ms co-located) and converges once the global lock releases. Sticky routing mitigates cross-replica staleness.

\begin{table}[t]
\centering
\footnotesize
\caption{Fail-closed verification scenarios.}
\label{tab:fault-scenarios}
\rowcolors{2}{rowgray}{white}
\begin{tabular}{@{}lp{5cm}@{}}
\toprule
\rowcolor{headergray}
Category & Failure Scenarios \\
\midrule
Network & 100\% packet loss, 500ms+ latency, connection reset, refused, DNS failure \\
Protocol & Malformed JSON, HTTP 4xx/5xx, empty body, truncated response \\
Agent & SIGKILL, SIGSTOP, slow startup, out-of-memory \\
Configuration & Invalid URL, wrong port, TLS mismatch \\
Overload & Agent at 150\% capacity \\
\bottomrule
\end{tabular}
\end{table}

To verify fail-closed behavior, we inject failures via \texttt{toxiproxy} between the database
and Cedar agent. Table~\ref{tab:fault-scenarios} catalogs the
scenarios tested across five categories. In all of them, the plugin returns \textsc{Ignore} and the native denial is preserved, \ie, we observe zero fail-closed violations.
For grant preservation, we compared
authorization decisions between baseline and Cedar-enabled MySQL
across a comprehensive test suite, checking that every request the baseline allows is also allowed by the Cedar-enabled build. Here too, we did not observe any violations.

\subsection{Residual risks}
\label{subsec:residual-risks}
We make explicit the residual risks that persist even when the mechanisms above function correctly.

\textbf{TCB compromise.} Attacks that compromise the database kernel, the plugin, or the Cedar agent fall outside our threat model (Sub-section~\ref{sec:threat-model}). Defending against such adversaries requires system-level isolation, \ie, hardened deployments, OS-level controls, or trusted execution environments, which is beyond our contribution.

\textbf{Timing side-channels.} Post-native authorization introduces differential timing, since external evaluation on a cache miss takes longer than a native decision. An observer can thus infer which requests traverse the external path and derive the coarse policy structure. Eliminating this channel would require latency equalization at non-trivial performance cost, which we do not pursue.

\textbf{Staleness under policy updates.} Authorization caching and deployments that relax synchronous invalidation for higher availability can create a bounded window during which stale decisions may be served. \sysname{} limits this window through TTLs and synchronous cache invalidation on policy and entity updates (Sub-section~\ref{subsec:entity-sync}).

\textbf{Semantic misconfiguration.} Schema and policy validation reject structurally invalid configurations but cannot verify that a policy reflects the administrator's intent. This remains an operational risk best addressed through review workflows, staged rollouts, and elaborate audit logging.

\section{Related Work}
\label{sec:related}

This section review some of the related literature that compare closely with the work done in this paper.

\textbf{In-database access control mechanisms.}
The earliest relational database authorization, \ie, the \textsc{grant}/\textsc{revoke} model~\cite{griffiths1976authorization} and subsequent RBAC extensions~\cite{eisenbergSQL1999Formerly1999}, remains the dominant paradigm in open-source databases. Rizvi et al.~\cite{rizvi_extending_2004} formalized fine-grained access control via authorization views, distinguishing the Truman model (silent query rewriting) from the Non-Truman model (rejection of insufficiently privileged queries). PostgreSQL Row-Level Security~\cite{postgres-rls} and Oracle Virtual Private Database~\cite{VirtualPrivateDatabases} implement this paradigm. Paul et al.\ introduced ASQL~\cite{paulPosterASQLAttribute2022} and its successor ASQLPLUS~\cite{meena2023efficiently}, which extend SQL syntax with attribute constructs and embed enforcement into MySQL's parser and execution engine. Both require deep, version-specific DBMS modifications. The \texttt{sepgsql} module integrates PostgreSQL with SELinux mandatory access control via security labels~\cite{F40SepgsqlSELinux2026}, but cannot enforce row-level policies and is limited to Linux with SELinux enabled.

Oracle Real Application Security (RAS)~\cite{oracle-ras} is the closest commercial analog. It supports application-managed ABAC sessions with fine-grained row and column privileges. However, RAS is proprietary, Oracle-specific, and its policy evaluation carries no formal verification guarantees. \sysname{} targets open-source databases, provides cross-DBMS uniformity, and grounds its PDP in Cedar's mechanically verified evaluator. Mohamed et al.~\cite{mohamedSystematicLiteratureReview2023} survey authorization across database models and identify fine-grained, context-aware access control as persistently unmet.

\textbf{Proxy-based and application-layer enforcement.}
Ferraiolo et al.~\cite{ferraiolo_imposing_2017} proposed NDAC, which applies the graph-based policy representation of Next Generation Access Control to relational databases by intercepting SQL queries through an external access manager proxy. Felt et al.~\cite{felt_diesel_2011} enforce per-module least privilege in database-backed web applications via an application-layer proxy, although at the cost of significant performance overhead. Colombo and Ferrari~\cite{colombo_fine-grained_2016} enforce fine-grained access control in document-oriented NoSQL stores using query rewriting. Gupta et al.~\cite{gupta_attribute-based_2021} manage ABAC policies through a Flask application, which are first converted into a compact set of roles and then are propagated to MongoDB. In a subsequent work, the same authors enabled ABAC by intercepting MongoDB Wire Protocol messages at a proxy for field-level ABAC~\cite{gupta_enabling_2023}. These proxy architectures introduce an additional network hop, create a single point of failure, and more critically, can be bypassed by clients with direct database access.

\textbf{Policy languages and authorization engines.}
Cedar~\cite{cedar2024oopsla} provides formally verified, terminating ABAC evaluation and is 42.8$\times$--80.8$\times$ faster than OPA/Rego in authorization benchmarks. Its static validator~\cite{disselkoen_how_2024} rejects type-inconsistent and non-terminating policies before deployment, which is a guarantee that no prior database authorization system provides. Chakarov et al.~\cite{chakarov2025verified} describe how AWS rebuilt its authorization engine in Dafny, sustaining $10^9$ authorizations per second under formal verification. 
Pang et al.~\cite{pang2019zanzibar} (Zanzibar) pioneered centralized, externalized authorization at Google scale. 
XACML~\cite{oasis2013xacml} defines the canonical PEP/PDP reference architecture that \sysname{} aligns with, but XML-verbose policy loading degrades sharply beyond 100 policies~\cite{turkmen2008xacml}, making it ill-suited for per-query database enforcement.

\textbf{Fine-grained access control in data platforms.}
Recent systems have integrated fine-grained governance into large-scale data platforms. Of these, Lakeguard~\cite{grund_databricks_2025} enforces row-level filters and column masks in Databricks/Spark workloads, BigLake~\cite{BigLakeBigQuerysEvolution} applies uniform column- and row-level controls across cloud analytics engines via BigQuery's Read API, while Data Guard~\cite{tran_data_2025} enforces purpose-based access control at LinkedIn's Spark and Trino data warehouses using pre-compiled SQL views. These systems target cloud-native or data-warehouse environments and assume control over the execution engine. \sysname{} addresses the complementary challenge of retrofitting ABAC into existing, unmodified RDBMS deployments where replacing the database engine is not feasible.

\begin{table}[t]
\centering
\scriptsize
\caption{Qualitative comparison of database authorization approaches.}
\label{tab:comparison}
\rowcolors{2}{rowgray}{white}
\setlength{\tabcolsep}{2pt}
\begin{tabular}{@{}llccccc@{}}
\toprule
\rowcolor{headergray}
\textbf{System} & \textbf{Approach} & \textbf{ABAC} & \textbf{Ext.\ ctx} & \textbf{Cross-DB} & \textbf{Verified} & \textbf{Bypass-proof} \\
\midrule
RLS~\cite{postgres-rls} & Native & \xmark & Limited & \xmark & \xmark & \cmark \\
\texttt{sepgsql}~\cite{F40SepgsqlSELinux2026} & Native & \xmark & \xmark & \xmark & \xmark & \cmark \\
Oracle RAS~\cite{oracle-ras} & Proprietary & \cmark & \cmark & \xmark & \xmark & \cmark \\
ASQLPLUS~\cite{meena2023efficiently} & Core mod. & \cmark & \cmark & \xmark & \xmark & \cmark \\
NDAC~\cite{ferraiolo_imposing_2017} & Proxy & \cmark & \cmark & \cmark & \xmark & \xmark \\
Diesel~\cite{felt_diesel_2011} & Proxy & \xmark & \xmark & \xmark & \xmark & \xmark \\
Lakeguard~\cite{grund_databricks_2025} & Platform & \cmark & \cmark & \xmark & \xmark & \cmark \\
Data Guard~\cite{tran_data_2025} & Platform & \cmark & \cmark & \xmark & \xmark & \cmark \\
\midrule
\sysname{} & Hook & \cmark & \cmark & \cmark & \cmark & \cmark \\
\bottomrule
\end{tabular}
\end{table}



\section{Discussion}
\label{sec:discussion}
In this section, the impacts of our approach are discussed along with the implications of alternative design choices and future extensions.

\textbf{Adoption considerations.}  Decoupled authorization in
\sysname{} is most valuable when requirements exceed native DAC/RBAC, \ie, for
context-dependent policies, cross-database policy uniformity, static checking and
analyzable policies.
Since policy evaluation resides entirely in the Cedar agent, extending to a richer access-control model such as ReBAC, or adopting a future Cedar version, reduces to updating the agent and its entity schema, with no change to the database plugin or core.
Porting \sysname{} to a new source-available relational database management system
beyond MySQL and PostgreSQL requires only identifying a post-native interception
point and implementing the corresponding request-marshaling shim.
Multi-database and heterogeneous deployments can reuse or share the same Cedar
agent through namespace-prefixed entity schemas. Adding a new instance of the already supported
databases, \ie, MySQL and PostgreSQL, only requires introducing a new namespace in the Cedar schema and registering the corresponding prefix in the plugin configuration. While we used Cedar for the reasons put forward in Section~\ref{sec:intro}, other external policy engines can also be explored.

\textbf{Alternative design choice.} \sysname{} combines the native and Cedar layers disjunctively, as described in Sub-section~\ref{subsec:post-denial}. It may be argued that such a design does not allow a conjunctive combination of native access control and external authorization, in the style of restrictive mechanisms like row-level security that only narrow what the privilege system grants. However, we did not make that design decision in \sysname{} for two reasons. First of all,
native DAC and RBAC in databases are themselves disjunctive, \ie, access is granted if any
applicable privilege applies. If there is still a specific requirement for making conjunctive combination of native access decision and external authorization, then hook invocation has to be made for every access request. This would slow down the response time. Second, and more importantly, it will require duplicating all the existing native grants to the Cedar agent, thereby increasing administrative burden and consistency risks.

Note that a composition like that done in \sysname{} does not make a deployment more permissive. With reference to the illustrative scenario introduced earlier, consider a hospital in which the \textit{cardiologist} role holds a coarse grant of \texttt{SELECT} on all patient records, only because native RBAC cannot state anything finer without causing role explosion. On migrating to \sysname{}, the administrator revokes this grant and writes a Cedar policy that permits a cardiologist to read only the records of cardiology patients, during working hours and from the hospital intranet. The effective permission set therefore shrinks and the configuration returns to least privilege, while the other requirements that remain native are enforced as before. Deriving such replacement policies automatically from existing grants is the policy synthesis question raised below.

\textbf{Open research questions.}  Our work on \sysname{} suggests several future directions of research. \emph{Policy synthesis}: How to derive Cedar policies automatically from existing GRANT permissions and query logs, reducing migration barriers. \emph{Caching formalization}: What level of consistency
guarantees can be maintained while aggressively caching authorization decisions, with provably bounded staleness. \emph{Compliance automation}: Cedar's formal verification capabilities could in principle be used to mechanically check policies against regulatory schemas (GDPR, HIPAA, SOX), but bridging policy semantics and regulatory language remains an open problem.


\section{Conclusion}
\label{sec:conclusion}

In this paper, we have proposed \sysname{}, which is a decoupled architecture that retrofits attribute-based
access control into a relational database without modifying its core. It enables the community to take a fresh look at database authorization mechanisms for supporting policies beyond their native DAC/RBAC capabilities. 
Our proof-of-concept implementations of \sysname{} for MySQL and PostgreSQL show acceptable performance and achieves the desired security goals. The detailed experimental results demonstrate that formal, context-aware authorization is practical for production database workloads. The code of all our implementation has been made open source for reproducibility and verifiability, and for enabling the database community to develop it further. We also feel that commercial closed source database vendors can consider providing support for such external authorization plugins.
%




\printbibliography

\end{document}